\documentclass[twocolumn]{pasj00}
\tenpoint

\usepackage{lscape}

\SetRunningHead{T. Miyakawa et al.}{Luminosity Dependepce of
the Electron Temperature}
\title{Luminosity dependence of the electron temperature \\ in the bright hard state of the black hole candidate GX 339--4}
\author{Takehiro \textsc{Miyakawa}, \altaffilmark{1,2} Kazutaka
\textsc{Yamaoka}, \altaffilmark{3} Jeroen \textsc{Homan},
\altaffilmark{4}  Koji \textsc{Saito}, \altaffilmark{5} \\ Tadayasu
\textsc{Dotani}, \altaffilmark{1}¡¡Atsumasa \textsc{Yoshida}, \altaffilmark{3} and
Hajime \textsc{Inoue}, \altaffilmark{1}}%
\altaffiltext{1}{The Institute of Space and Astronautical Science (ISAS), Japan
Aerospace Exporlation Agency (JAXA), \\ 3-1-1 Yoshinodai, Sagamihara, Kanagawa 229-8510}
\altaffiltext{2}{Department of Astronomy, Graduate School of Science,
the University of Tokyo,\\
7-3-1 Hongo, Bunkyo-ku, Tokyo 113-0033}
\email{miyakawa@isas.jaxa.jp}
\altaffiltext{3}{Department of Physics and Mathematics, Aoyama Gakuin
University, \\ 5-10-1 Fuchinobe, Sagamihara, Kanagawa 229-8558}
\email{yamaoka@phys.aoyama.ac.jp}
\altaffiltext{4}{MIT Kavli Institute for Astrophysics and Space Research, 70 Vassar Street, Cambridge, MA 02139, USA}
\altaffiltext{5}{Department of Physics, Tokai University, 1117 Kitakaname, Hiratsuka, Kanagawa 259-1292}
\Received{$\langle$2007 April 27$\rangle$}
\Accepted{$\langle$2008 February 12$\rangle$}
\Published{$\langle$publication date$\rangle$}
\begin{document}
\maketitle

\begin{abstract} We have analyzed 200 Rossi X-ray Timing Explorer
 observations of the black hole candidate GX 339--4, all from the bright
hard state periods between 1996 and 2005.  Purpose of our study
 is to investigate the radiation  mechanisms in the hard state of  GX
339--4. The broadband 3--200 keV spectra were successfully modeled by a
simple analytic model, power--law with an exponential
cut-off modified with a smeared edge. The obtained energy cut-off  
($E_{\rm{cut}}$) was distributed over 50--200 keV, and the photon
index over 1.4--1.7.  We found a clear anti-correlation
($E_{\rm{cut}} \propto L^{-0.70\pm0.06}$)    between the X-ray
luminosity ($L$) in 2--200 keV and $E_{\rm{cut}}$ , when $L$  is
larger than $7 \times 10^{37}$ erg s$^{-1}$ (assuming a distance of 8
kpc), while $E_{\rm{cut}}$ is roughly constant at around 200 keV
when $L$ is smaller than $7 \times 10^{37}$ erg s$^{-1}$. This
anti-correlation remained unchanged by adopting a more physical 
thermal Comptonization model, which resulted in the anti-correlation
that can be expressed as $kT_{\rm{e}} \propto L^{-0.24\pm0.06}$.
These anti-correlations can be quantitatively explained by a picture
in which the energy-flow rate from protons to electrons balances with
the inverse Compton cooling.  \end{abstract}

\section{Introduction}

Black hole X-ray binaries show large varieties in their X-ray
properties. Several spectral and variability states have been
recognized in the past, with varying names and varying definitions
(see e.g.\ McClintock \& Remillard 2006, Homan \& Belloni 2005) and it
is believed that they correspond to different accretion geometries. In
this paper we focus on the spectral properties of the so-called hard
state, which is the state observed at low X-ray
luminosities (i.e.\ below a few percent of the Eddington luminosity,
$L_{\rm Edd}$), although it can also be seen during the rising phase
of transient outbursts below $\sim$0.2$L_{\rm Edd}$.
The accretion geometry in some of the
X-ray states is relatively well understood -- e.g.\ the soft state is
thought to be governed by an optically thick and geometrically thin
accretion disk -- this is not the case for the hard state. X--ray
spectra in the hard state are well represented by a power--law with a
photon index of 1.4$\sim$1.7  (Tanaka \& Shibazaki 1996) and, in some
sources, a high energy cut-off at $\sim$100 keV (Grove et al.\ 1998).
Several models have been proposed to explain the hard state spectra.
Broadband X-ray energy spectra of a large number of black hole binaries
in the hard state have been successfully modeled by thermal Comptonization
model (Dove et al.\ 1997, Pountanen \&  Svensson\ 1996), although 
location and geometry of the Comptonizing medium is still debated.
Other models include advection dominated accretion flows (ADAF) (Esin
et al.\ 1997), synchrotron and/or synchrotron self-Compton radiation
from the base of a jet (Markoff, Nowak, \& Wilms 2005), and external
Compton scattering in the jets (Georganopoulos, Aharonian, and Kirk
2002).

\begin{figure*}[htbp] \begin{center}
\FigureFile(61mm,50mm){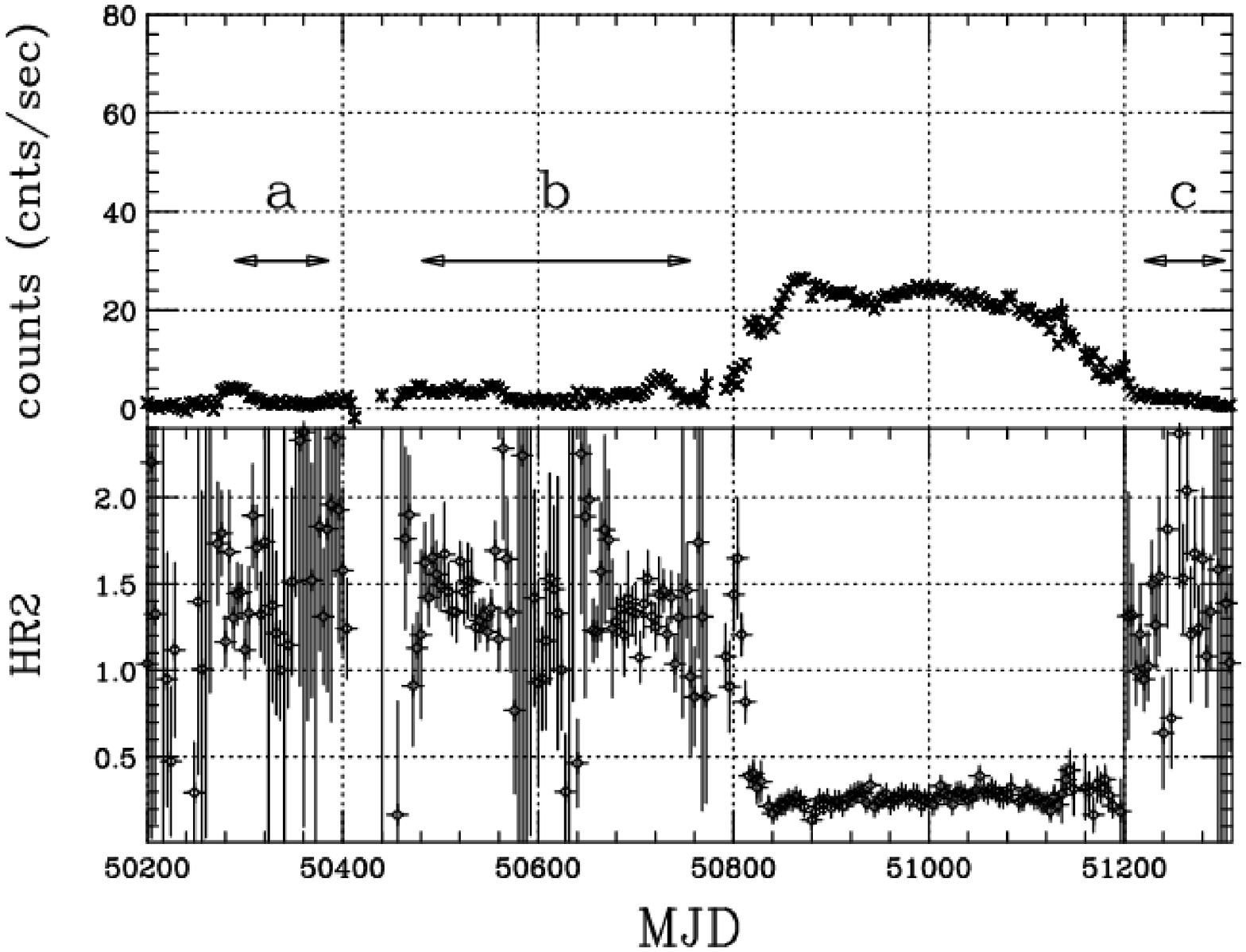} \hspace{0.0cm}
\FigureFile(57mm,50mm){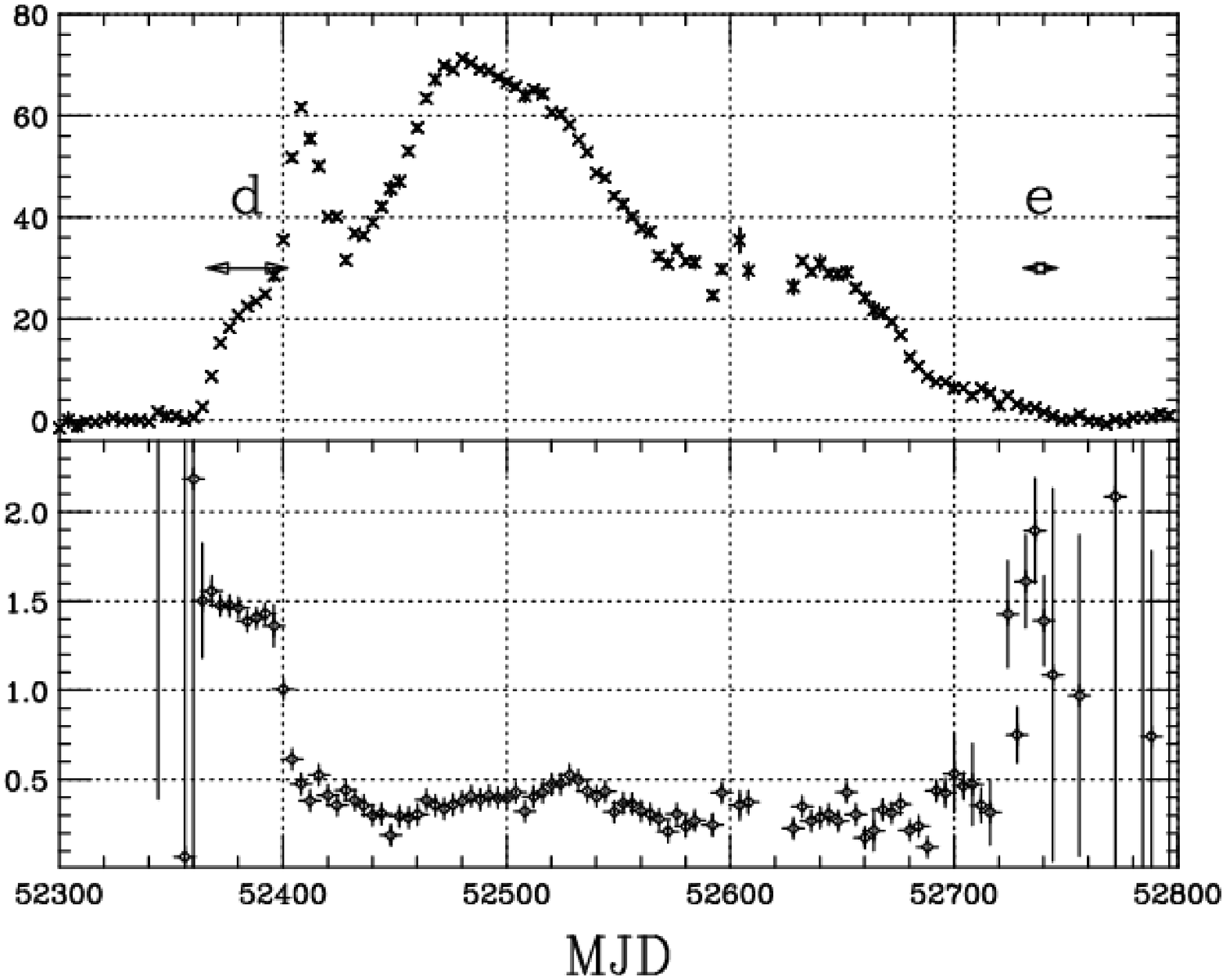} \hspace{-0.2cm}
\FigureFile(57mm,50mm){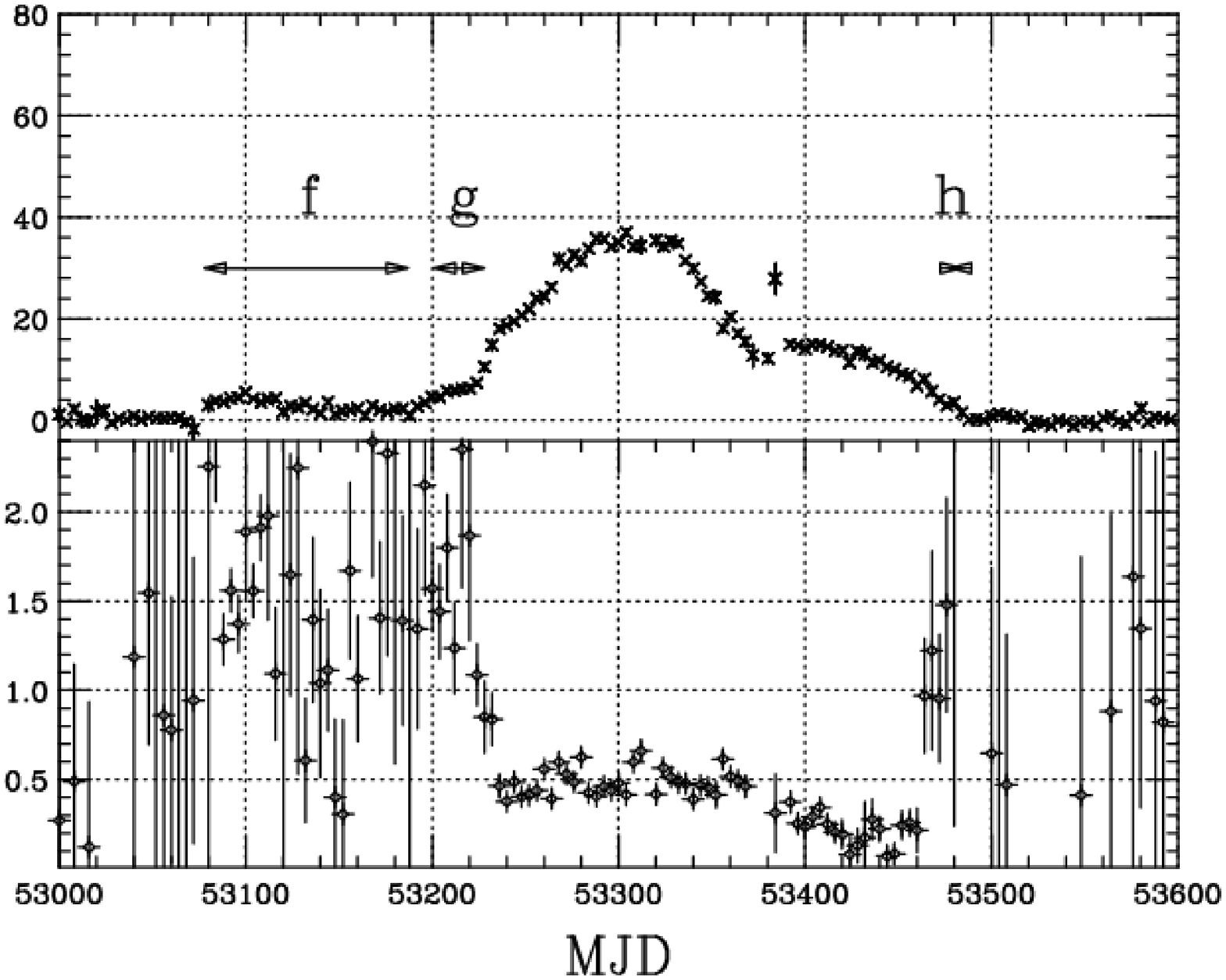} \end{center}
\caption{RXTE/ASM light curves (upper panels) and hardness ratios HR2
(= 5--12 keV/3--5 keV; lower panels) of the black hole candidate GX
 339--4 in the active periods from 1996 to
2005. The arrows labelled $a$ to $h$ indicate the hard state
intervals of which the RXTE/PCA observations were analyzed in this paper.}
\label{lc_1996_2005} \end{figure*}

GX 339--4 was discovered with the X-ray satellite OSO-7 in 1971
(Markert et al.\ 1973). Since it was similar to Cygnus X--1 in terms
of X-ray spectral and variability properties, it was considered
a black hole candidate (BHC, Samimi et al.\ 1979). A recent
measurement of the mass function (5.8$\pm$0.5 $M_\odot$, Hynes et al.
2004) strengthens this. We thus assume a mass of 5.8 $M_\odot$ for
this  black hole candidate as a secure lower limit. GX 339--4 is one
of the best-studied BHCs in X-rays and gamma-rays; it was observed
with Ginga/LAC (Ueda, Ebisawa, and Done 1994), CGRO/OSSE (Grabelsky
et al.\ 1995, Smith et al.\ 1999), ASCA  (Wilms et al.\ 1998), RXTE
(Smith et al.\ 1999, Belloni et al. 2005), Beppo-SAX  (Corongiu et
al.\ 2003) and INTEGRAL (Belloni et al.\ 2006, Joinet et al.\ 2006). 
GX 339--4 has shown  various spectral states during its pre-RXTE
outbursts (Tanaka \& Shibazaki 1996), but often stayed in the
low-luminosity hard state (low/hard state). All the spectra in the
low/hard state can be roughly described by a Comptonization model,
requiring the addition of a reflection component, soft excess and
iron-K lines. Detailed broadband analysis in the 2--1000 keV range
was made using Ginga, RXTE and OSSE data from 1991 and 1996
observations by Zdziarski et al.\  (1998) and Wardzi$\acute{n}$ski et
al.\ (2002).  Wardzi$\acute{n}$ski et al (2002) concluded that the
four hard state spectra have very similar intrinsic photon indices of
$\cong$ 1.75. Furthermore, they found that the energy cut-off 
possibly decreases with increasing the luminosity. Zdziarski et al.\ 
(2004) studied the long-term behavior by compiling GINGA/ASM,
CGRO/BATSE and RXTE/ASM data (spanning 16 years) and also found that
the electron temperature depends on the luminosity by showing a
positive correlation between the BATSE flux and the photon index in the
70--160 keV range which reflects the breaking energy.

GX 339--4 has also been extensively studied at other wavelengths. The
optical counter part was first identified by Doxsey et al.\ (1979) as
a V$\sim$18 blue star. However, even during the X-ray off state,
optical spectroscopic observations with the VLT revealed no spectral
features from the  companion star (Shahbaz et al.\ 2001). Hence, the
type and distance of the secondary star are still unknown. There are
two papers which give a lower limit on the distance: Maccarone
(2003), 7.6 kpc,  and Hynes et al.\ (2004), 5.6 kpc. A careful study
of the distance is presented by Zdziarski et al.\ (2004) who argue
for a most likely distance of 8 kpc. We assume a distance of 8 kpc in
this paper. Multi-wavelength observations have shown that the radio
emission of GX 339--4 correlates very tightly with the X-ray fluxes over
more than two orders of magnitude, suggesting that the
jet plasma may also play a role in high energy band (Corbel et al .\
2000, Corbel et al.\ 2003,  Markoff et al.\ 2003). Corbel et al.\
(2003) showed that a significant  fraction of the X-ray flux observed
in the low/hard state of black hole  candidates may be due to
optically thin synchrotron emission from the  compact jet. Also 
jet emission have been found in the near-infrared 
(Corbel \& Fender 2002) and Homan et al. (2005) discovered a tight
relation between near-infrared and X-ray fluxes in the hard state of
GX 339--4, similar to that of the X-ray/radio correlation.

Many authors have only studied a small number of observations during
individual outbursts of GX 339--4.  In order to improve our
understanding of the  radiation mechanisms in the hard state of GX
339--4,  we have performed a systematic study of detailed
correlations among spectral parameters, using the large archive of
public RXTE data for GX 339--4.  In Section 2 we describe our data
analysis, and in Section 3 we present our results, focusing on the
correlations between spectral parameters such as luminosity and high
energy cut-off or electron temperature. Finally in Section 4, we will
discuss the origin of the observed correlations.

\section{Observation and Data Reduction}

GX 339--4 has been observed for over 10 years with the Rossi X-ray
Timing Explorer (RXTE, Bradt et al. 1993). RXTE carries three
scientific instruments: the Proportional Counter Array  (PCA: Jahoda
et al.\ 1996), the High Energy X--ray Timing Experiment (HEXTE; 
Rothschild et al.\ 1998),  and the All-Sky Monitor (ASM: Levine et
al.\ 1996).  

Figure $\ref{lc_1996_2005}$ shows the 1.5--12 keV RXTE/ASM light
curves of GX 339--4 from three active periods in the
last 10 years: in 1996--1999, 2002--2003, and 2004--2005 (data from
the 2007 outburst were not analyzed). Following McClintock $\&$
Remillard (2006), we also plot the ASM hardness of HR2 (defined
as the  ratio of ASM counts in 5--12 keV to counts in 3--5 keV),
where values of $\sim$1.5 are typical in the hard state. 

Selection of the RXTE/PCA observations for our analysis was based on
the location of these observations in a hardness-intensity diagram,
as shown in Figure $\ref{selection}$. This diagram was created from
 the background subtracted Standard 2 mode PCA data (see below), where 
count rates are taken from the 3.5--20 keV band and the hardness defined as
the ratio of count rates in the 10--20 keV and 3.5--5.5 keV bands.
All count rates were normalized to those of the Crab, to account for
possible changes in the detector response. Hard states correspond to the
slightly curved vertical branches in the right part of the diagram.
For selecting hard state observations for our analysis we used two
criteria: hardness $>$1.4 (to select hard states) and  count rate $>$
15 mCrab (to get enough counts per spectrum). Thus, we obtained 200
RXTE observations. The periods during which these observations were
made are indicated by the horizontal arrows in Figure
$\ref{lc_1996_2005}$.  A list of all the selected observation IDs is
given in Tables $\ref{table_1}$--$\ref{table_4}$.

\begin{figure}[hbtp]
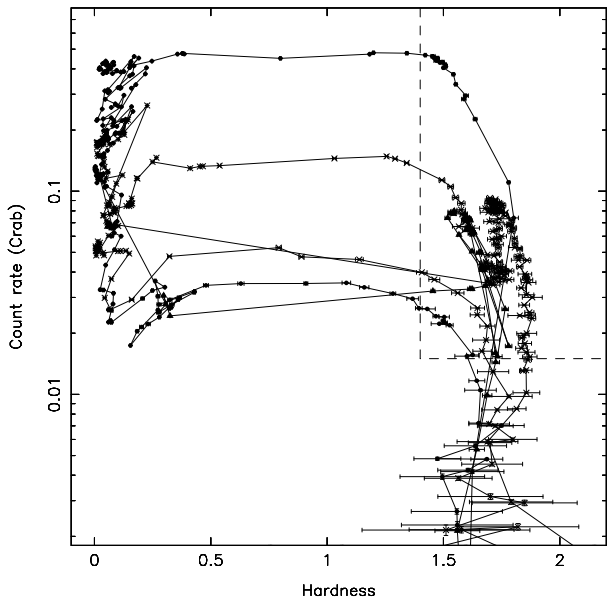
 \begin{center}
\FigureFile(80mm,50mm){figure2.ps} \end{center}
\caption{Hardness-Intensity diagram of the 1996/1997, 2002/2003 and
2004/2005 outbursts, created from RXTE/PCA data. The area in the upper
 right (outlined with the dashed lines) indicates the selected observations.}
\label{selection} \end{figure}

For the spectral data reduction we used the HEADAS 6.0.2 software
package provided by NASA/GSFC. Additional selection criteria for both
PCA and HEXTE are as follows: 1) elevation angle from the Earth
was larger than 10 degree, and 2) the offset angle was smaller than
0.02 degree. PCA spectra were only extracted from Proportional Counter
Unit 2 (PCU2), since it was the only PCU that was active in all
the selected observations. We used Standard 2 mode data, which
has a time resolution of 16 sec and 129 energy channels covering the
full PCA ($\sim$2--60 keV) energy range. The spectra were deadtime
corrected (a few \%), and the background, estimated by the background
model for bright sources, was subtracted.  As for HEXTE, we used the
archive mode data with a time resolution  of  16 sec taken from both
cluster A and B, and subtracted the background taken from the rocking
motion.  Deadtime correction for extracted spectra was made using
{\it hxtdead}.

We added 2\% systematic errors to each PCA spectral bin so as to
obtain  ${{\chi}_{\nu}}^2 \sim 1$ for the spectrum of the Crab. The
PCA energy range was limited to 3--20 keV and the HEXTE energy range
to 18--200 keV. Our PCA and HEXTE response files gave consistent
values of the photon index of the Crab spectrum for the PCA
(2.08$\pm$0.02) and HEXTE (2.10$\pm$0.02), indicating that the PCA and
HEXTE could be fitted simultaneously. We did so by allowing the
normalization of the HEXTE spectrum relative to that of the PCA
spectrum to vary. We found that the normalization factor varied
between 0.90--0.95. 

Two simple spectral models were applied to all the data in carrying
out the systematic spectral analysis of the continuum. One is a
simple analytic model, a power-law with an exponential cut-off
(cut-off power-law  in XSPEC, hereafter CPL), and the other is a
thermal Comptonization model with a spherical geometry, introduced by
Sunyaev \& Titarchuk (1980) (compst in XSPEC, hereafter COMPST).  In
addition to their normalization parameters, both models are described
by two other parameters:  photon index ($\alpha$) and energy of
spectrum cut-off ($E_{\rm{cut}}$) in the former model, and  electron 
temperature ($kT_{\rm{e}}$) and  Thomson optical depth ($\tau$) of a
high temperature plasma in the latter. These continuum models were
further modified by galactic absorption (wabs in XSPEC) and a
smeared edge (Ebisawa et al.\ 1994). The hydrogen column density in
the wabs model was fixed to $5 \times 10^{21}$ cm$^{-2}$ (Ilovaisky
et al.\ 1986). Edge energy and width in the smeared edge model 
were fixed at 7.11 keV (corresponding to the
neutral iron-K edge) and 10 keV, respectively. 

Absorption-corrected X-ray luminosities ($L$) were calculated in the 
2--200 keV range and were based on the PCA fit parameters. We assumed 
the distance of GX 339--4  to be 8 kpc (Zdziarski et al.\ 2004).  The obtained
luminosities ranged from 1.0$\times 10^{37}$ erg ${\rm{s}}^{-1}$ to 
2.1$\times 10^{38}$ erg ${\rm{s}}^{-1}$, corresponding to 1.3--29$\%$ 
of the Eddignton limit ($L_{\rm Edd}$=7.3$\times 10^{38}$ erg
${\rm{s}}^{-1}$), for a black hole mass of 5.8$M_{\odot}$. 

\section{Analysis and Results}

\begin{figure}[t]
\begin{center}
\FigureFile(80mm,50mm){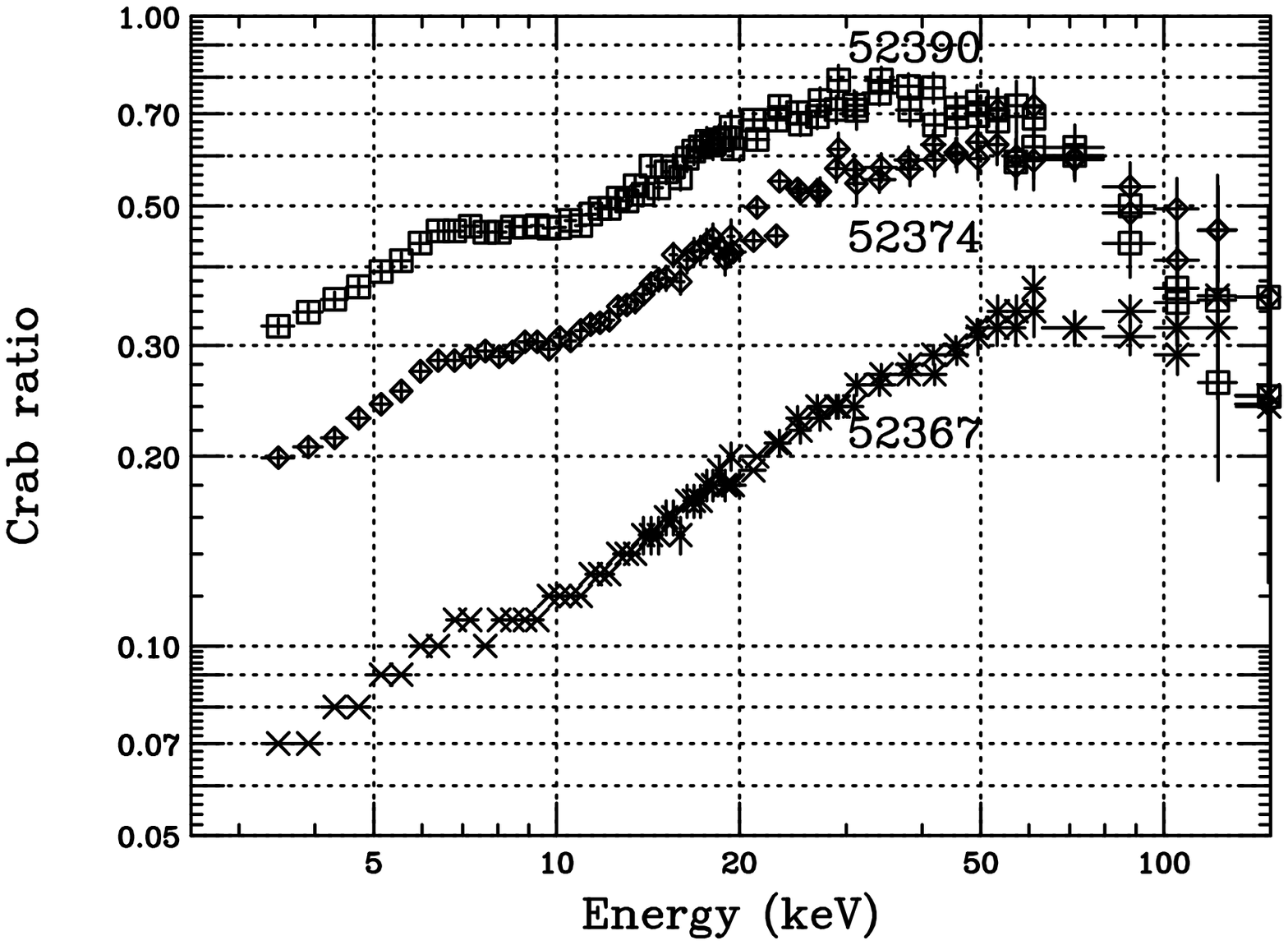}
\end{center}
\caption{PHA ratio of GX 339--4 to Crab taken during different
 epochs. The high energy cut-off depends on the X-ray luminosity, while the
 slope of this spectrum remains unchanged. }
\label{ratio_spectrum}
\end{figure}

To show the presence of the high energy cut-off in GX 339-4 in a
model independent manner, we calculated the PHA ratio of GX 339-4 to the
Crab Nebula for several observations. A Crab spectrum from 2002 April
28 was extracted in the same way as described in the previous
Section.  This spectrum could be fit in the 3--200 keV range by a
simple featureless  power-law model with a photon index of
2.10$\pm$0.01. GX 339-4 spectra were taken from  observations on
April 3rd 2002 (MJD 52367), April 10th 2002 (MJD 52374), and April
26th 2002 (MJD 52390). Figure $\ref{ratio_spectrum}$ shows the ratio
of the three GX 339--4 spectra to the Crab spectrum.   As can be seen
from this figure, there is a clear cut-off at high energies that
changed from $\sim$30 keV to $\sim$70 keV with decreasing flux, while
the spectral slope, i.e. photon index, does not seem to vary
significantly. 

\begin{figure*}[htbp]
\begin{center}
\FigureFile(80mm,50mm){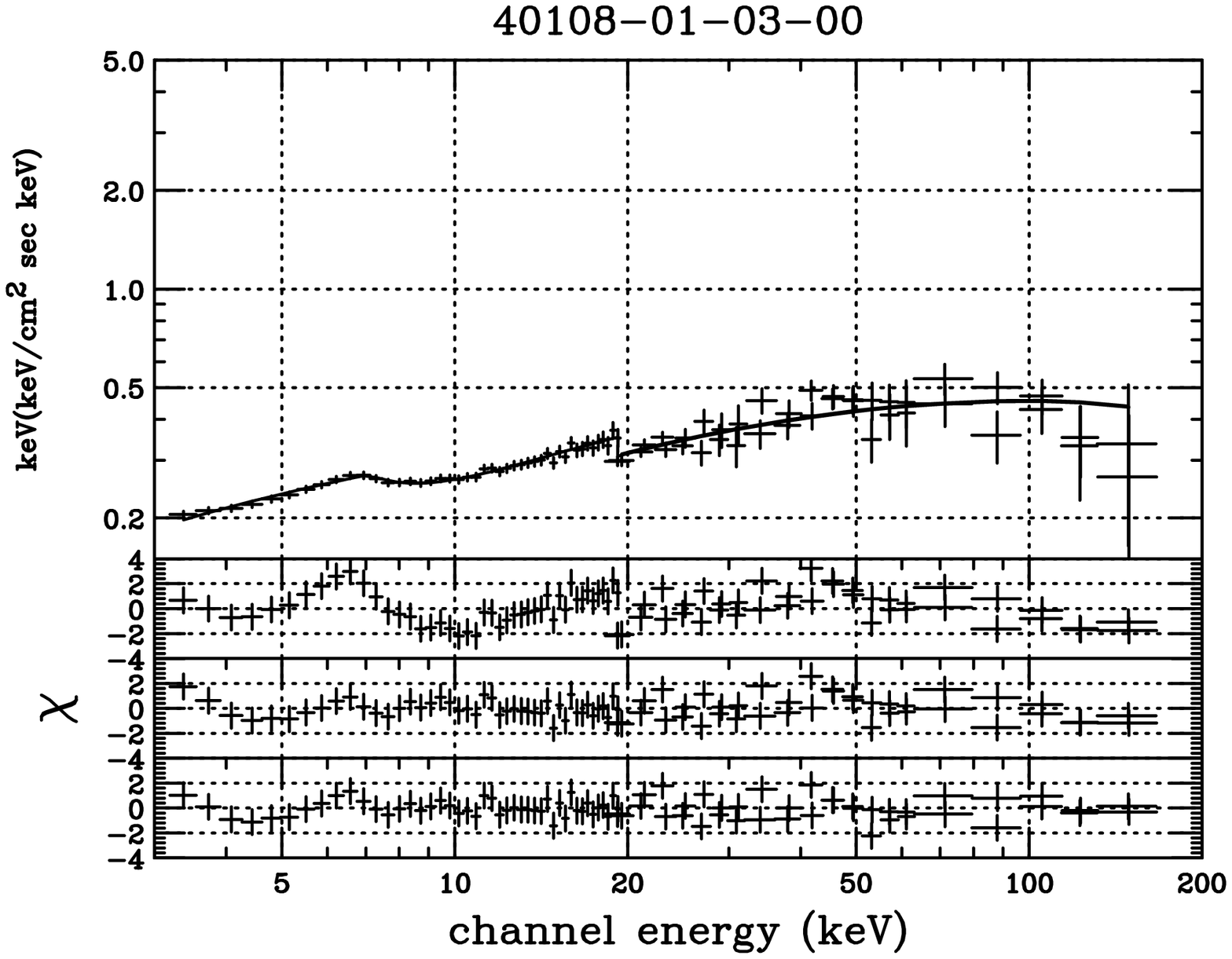}
\FigureFile(80mm,50mm){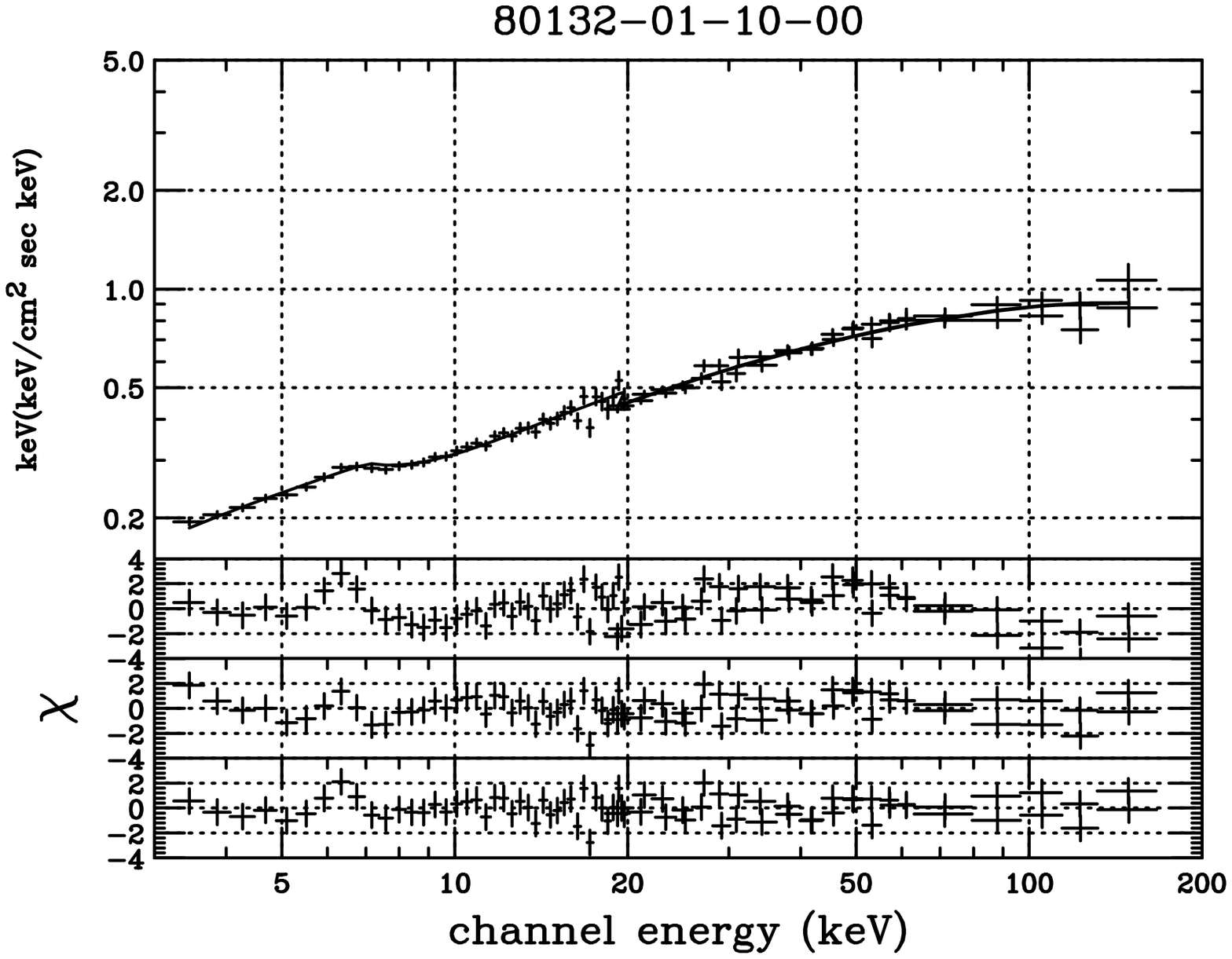}
\FigureFile(80mm,50mm){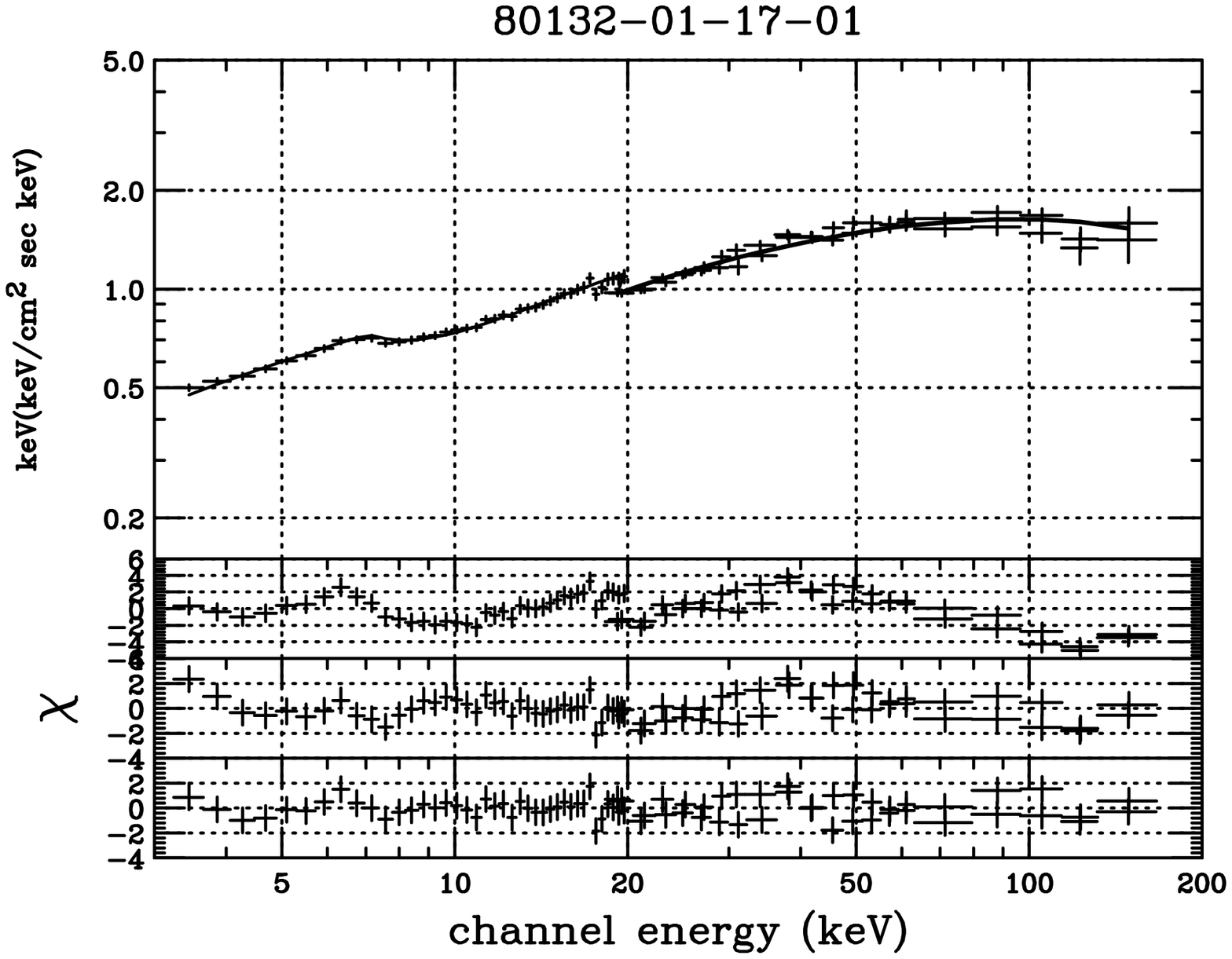}
\FigureFile(80mm,50mm){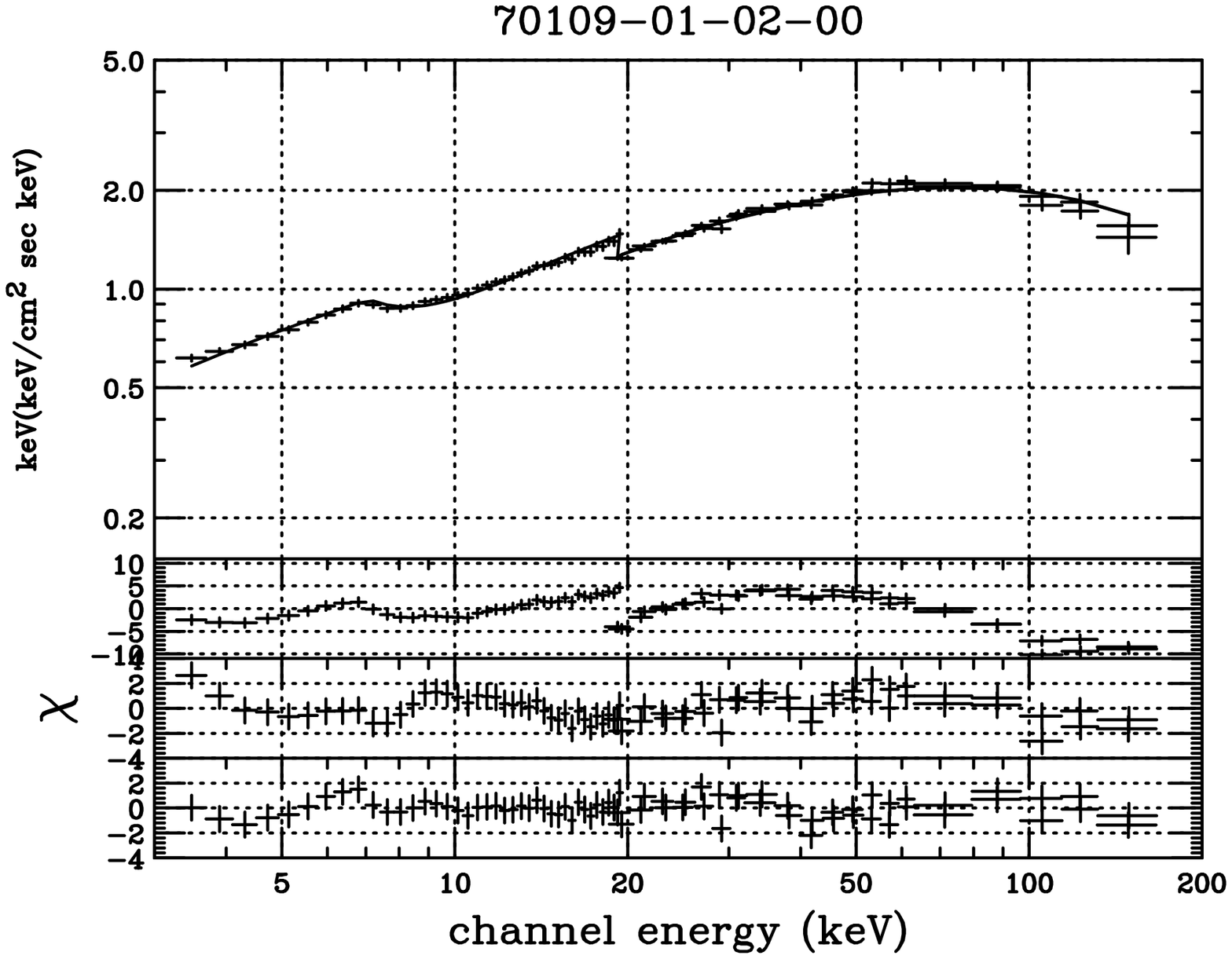}
\FigureFile(80mm,50mm){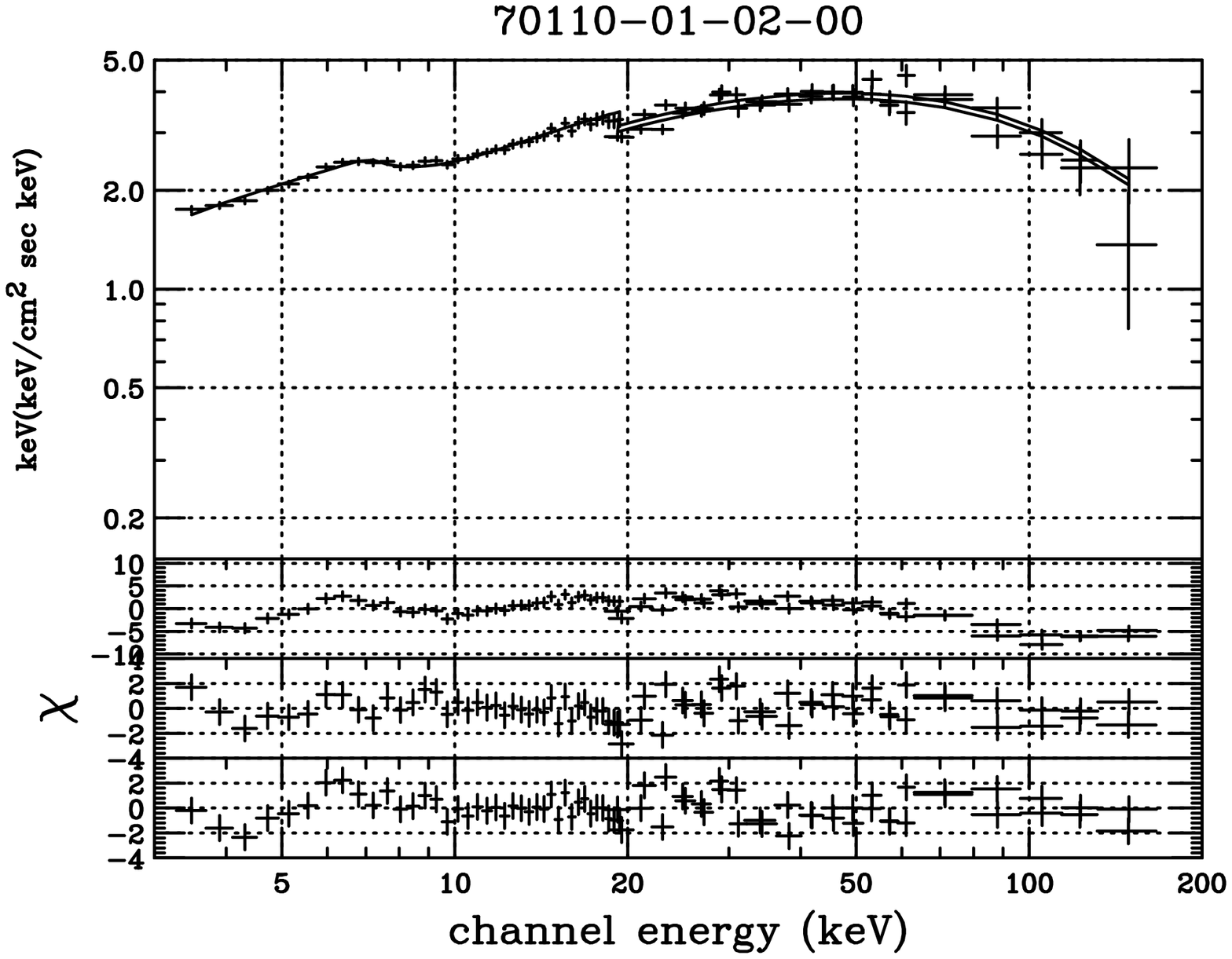}
\end{center}
\caption{Spectral fit results of the combined PCA and HEXTE spectra of GX
 339--4 for 5 representative observations.  
The top panels show an unfolded $\nu F_{\nu}$ spectrum fitted with a cut-off power-law model, 
while the three bottom panels show residuals with a simple power law,
 cutoff power-law, and thermal Comptonization (COMPST) model, respectively. } \label{spectrum_cut-off_Compton}
\end{figure*}

Next, we performed a systematic study of the spectra with the two
models described in Section 2. In the CPL model fitting,  we set
the maximum value at 500 keV for the energy cut-off $E_{\rm{cut}}$. 
The average and standard deviations  of the reduced chi-squares 
(${{\chi}_{\nu}}^2$ $\pm$ ${\sigma}_{\chi^2}$, for 74 degrees of
freedom) were 1.0$\pm$0.2 and 1.0$\pm$0.3 for the CPL and COMPST
models, respectively. Figure \ref{spectrum_cut-off_Compton} shows
example fits to  broadband spectra of selected five observations.
Both models could fit the 3--200 keV spectra for
each observation reasonably well.   The effects of a possible soft
excess and the iron-line features at 6.4 keV were negligible and
${{\chi}_{\nu}}^2$ did not improve  significantly when we added a disk
blackbody or a Gaussian component to our fit model. However, in cases
when $kT_{\rm{e}}$ was smaller than 20 keV, we found that it was
difficult to fit the spectrum with  the COMPST model, resulting in
${{\chi}_{\nu}}^2$ values that exceeded 2 (as we show later, 
reflection models were required in those cases).  Fit parameters in 5 
representative observations are given in Table $\ref{various_fit}$, 
and all the fit parameters are given in Tables  
$\ref{table_1}$--$\ref{table_4}$.

\begin{longtable}{lccccccc}
\caption{Best-fit parameters for fits with various
   model.}\label{various_fit}
\endhead
\endfoot
\multicolumn{7}{@{}l@{}}{\hbox to 0pt{\parbox{85mm}{\footnotesize
Notes. Errors are quoted at statistical 90\% level.
    }\hss}}
\endlastfoot
\hline\hline
Observation ID  & MODEL & $L$(10$^{37}$ erg s$^{-1}$)   &  $E_{\rm
   cut}$ /  k$T_{\rm e}$ (keV)  &  $\alpha$ / $\tau$  &  $\Omega$/2$\pi$ & $\chi^2$/d.o.f\\\hline
40108-01-03-00 & power-law  &  2.2$\pm$0.1  &  --  &  1.70$\pm$0.02  &  &   82/80 \\
               & CPL     &  2.0$\pm$0.1  &  $250_{-69}^{+144}$  &   1.62$\pm$0.04 &  & 60/79 \\
               & CPL+refl&  2.0$\pm$0.1  &  $>$500              &   1.83$\pm$0.03 &0.33$_{-0.10}^{+0.12}$  & 76/78 \\              
               & COMPST  &  1.9$\pm$0.1  &  27$\pm$3  &  $4.18_{-0.28}^{+0.26}$  & &  52/79  \\ 
               & COMPST+refl& 1.9$\pm$0.1 & 46$_{-11}^{+29}$ & 2.75$_{-1.07}^{+0.56}$ &  0.26$\pm$0.13 & 72/78  \\              
               & COMPPS     & 1.8$\pm$0.3 & 89$_{-23}^{+22}$ & 1.53$_{-0.35}^{+0.60}$ & 0.19$_{-0.09}^{+0.11}$ & 87/78 \\
80132-01-10-00 & power-law  &  3.1$\pm$0.1  &  --  &  1.54$\pm$0.01  &  &  143/75 \\
               & CPL     &  3.1$\pm$0.2  &  $234_{-41}^{+61}$  &  1.41$\pm$0.03  & & 75/74 \\
               & CPL+refl&  3.1$\pm$0.1  &  $>$500             & 1.61$_{-0.04}^{+0.02}$ & 0.33$_{-0.10}^{+0.03}$  & 79/73 \\              
               & COMPST  &  2.8$\pm$0.1  &  30$\pm$2         &  $4.46_{-0.18}^{+0.17}$  & &   98/74  \\ 
               & COMPST+refl& 3.0$\pm$0.1  & 40$_{-6}^{+9}$  &  3.60$_{-0.42}^{+0.36}$ & 0.28$\pm$0.10 & 59/73 \\              
               & COMPPS     & 3.0$\pm$0.2  & 66$_{-2}^{+16}$ &  $<$2.42 & 0.14$_{-0.05}^{+0.06}$ & 69/73 \\
80132-01-17-01 & power-law  &  6.6$\pm$0.1  &  --  &  1.62$\pm$0.01  &  &  249/75 \\
               & CPL     &  6.5$_{-0.2}^{+0.3}$    &  $168_{-20}^{+26}$  &  1.44$\pm$0.02  & & 72/74 \\
               & CPL+refl&  6.7$\pm$0.2     &  $>$500  & 1.70$_{-0.03}^{+0.02}$ & 0.42$_{-0.04}^{+0.08}$  & 104/73  \\              
               & COMPST  &  6.0$\pm$0.2    &  27$\pm$2  &  4.57$\pm$0.14 & &  90/74  \\
               & COMPST+refl & 6.2$\pm$1.1 & 36$_{-3}^{+5}$ & 3.61$_{-0.30}^{+0.28}$ & 0.27$_{-0.08}^{+0.04}$ & 60/73 \\
               & COMPPS      & 6.2$\pm$0.3 & 58$_{-1}^{+12}$ & $<$2.49 & 0.18$_{-0.05}^{+0.03}$ &  86/73\\
70109-01-02-00 & power-law  &  8.1$\pm$0.1  &  --  &  1.64$\pm$0.01  & &   862/74 \\
               & CPL     &  $8.4_{-0.2}^{+0.3}$  &  $115_{-7}^{+8}$  &  1.36$\pm$0.02  & & 80/73 \\
               & CPL+refl&  8.4$\pm$0.3    &  $>$470            & 1.64$_{-0.03}^{+0.04}$ & 0.46$_{-0.08}^{+0.09}$   & 201/72 \\              
               & COMPST  &  7.4$\pm$0.2     &          25$\pm$1  &  4.76$\pm$0.09  &  & 141/73  \\
               & COMPST+refl & 7.4$\pm$0.3   & 25$\pm$1          & 4.74$_{-0.09}^{+0.10}$ & 0.33$\pm$0.07 & 127/72 \\
               & COMPPS      & 8.1$\pm$0.1   & 58$_{-1}^{+2}$ & $<$2.92 & 0.33$_{-0.04}^{+0.05}$ & 100/72 \\
70110-01-02-00 & power-law  &  17.8$\pm$0.3  &  --  &   $1.76_{-0.02}^{+0.01}$   & &   534/74 \\
               & CPL     &  16.8$\pm$0.8     &  $80_{-7}^{+8}$  &   1.44$\pm$0.03  & &  82/73 \\
               & CPL+refl&  16.8$\pm$0.7     &  212$_{-40}^{+61}$ & 1.72$\pm$0.05  & 0.41$_{-0.11}^{+0.13}$ & 115/72 \\              
               & COMPST  &  16.1$\pm$0.8     &  20$\pm$1  &  5.02$\pm$0.16  & &  126/73  \\
               & COMPST+refl&16.3$\pm$0.4    &  25$_{-2}^{+3}$  & 4.12$_{-0.32}^{+0.30}$ & 0.31$_{-0.12}^{+0.13}$ &  81/72 \\              
               & COMPPS     &16.7$\pm$0.5    & 47$_{-2}^{+3}$   & $<$2.81 & 0.44$\pm$0.07 & 86/72 \\\hline
\end{longtable}

\begin{figure}[htbp]
\begin{center}
\FigureFile(80mm,80mm){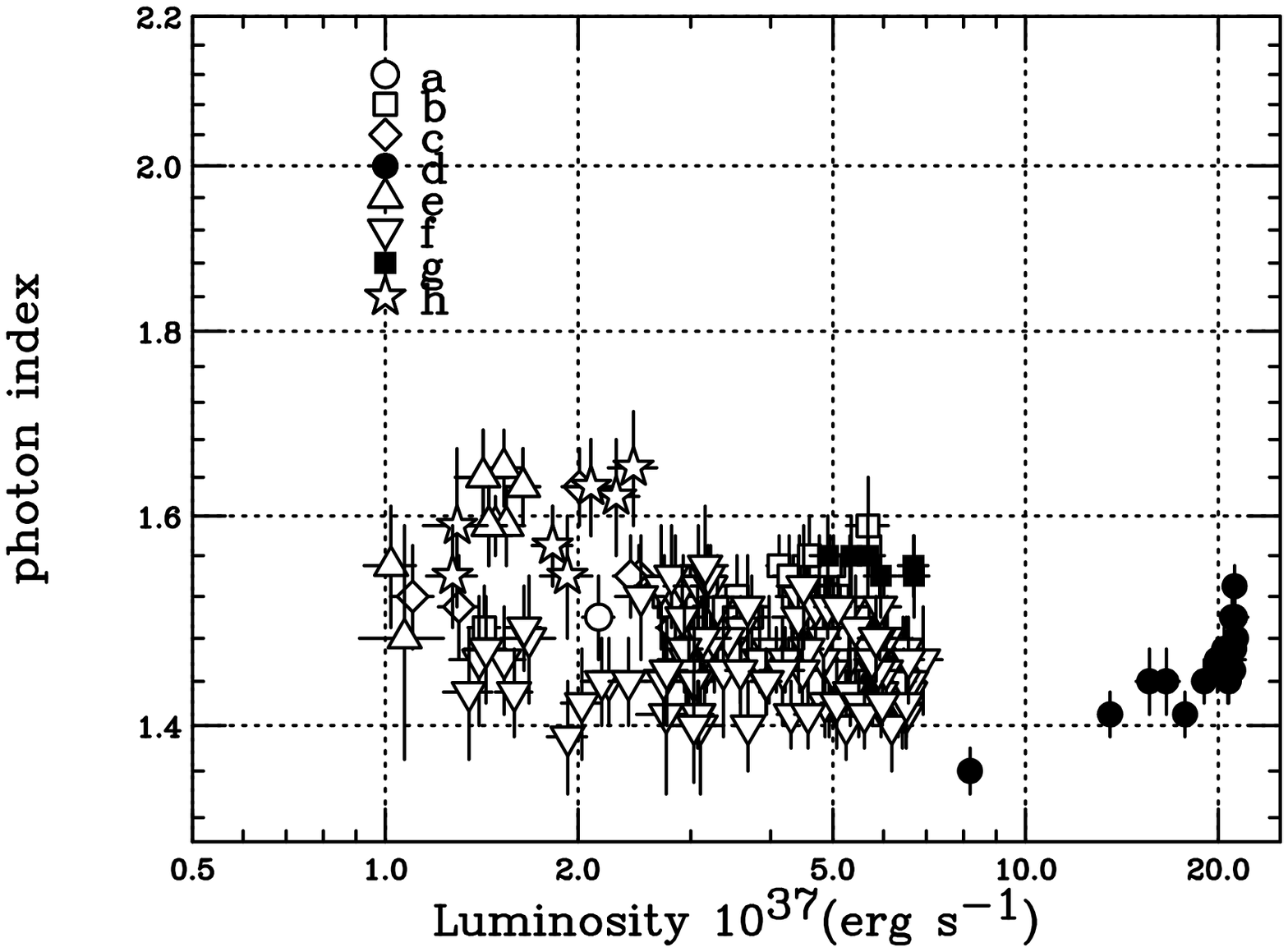}
\FigureFile(80mm,80mm){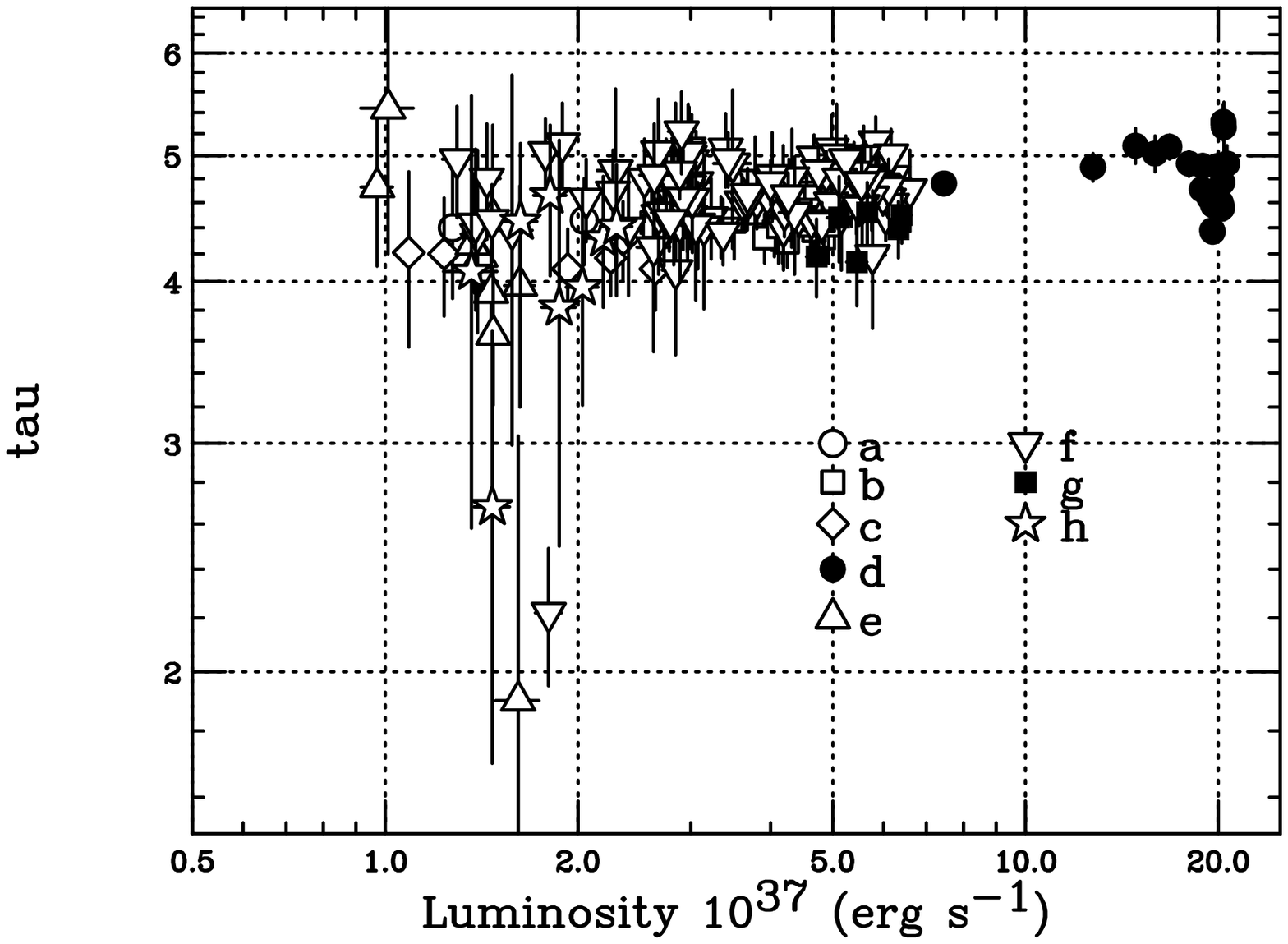}
\end{center}
\caption{Top: Relation between the luminosity in 2-200 keV($L$) and the photon
 index($\alpha$) obtained by the cut-off power-law model. Bottom:
 relation between the luminosity($L$) and Thomson optical depth($\tau$) obtained
 by COMPST model. The different symbols correspond to the epoch defined
 in Figure \ref{lc_1996_2005}. }
\label{L_pho_tau}
\end{figure}

\begin{figure}[htbp] \begin{center}
\FigureFile(80mm,80mm){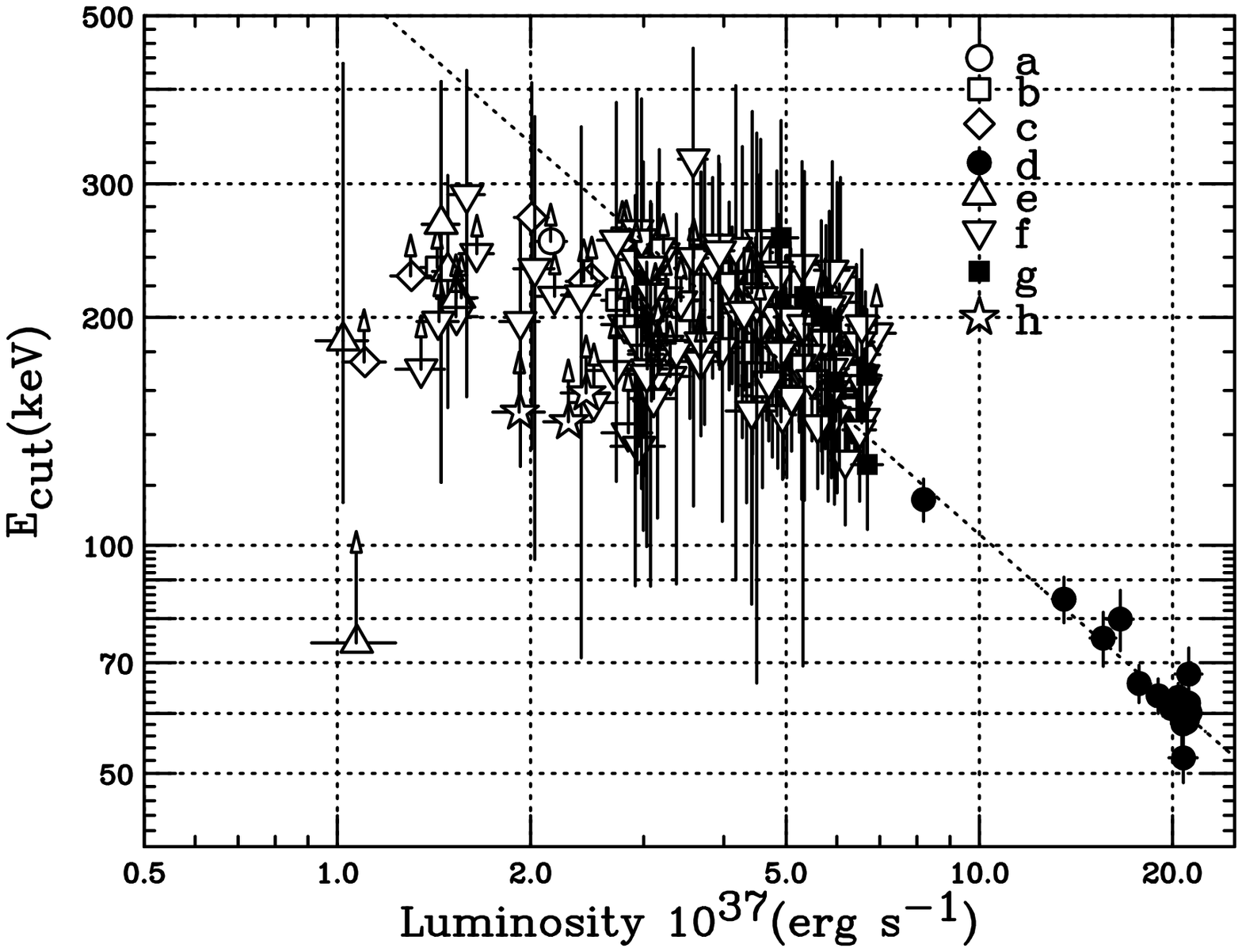}
\FigureFile(80mm,80mm){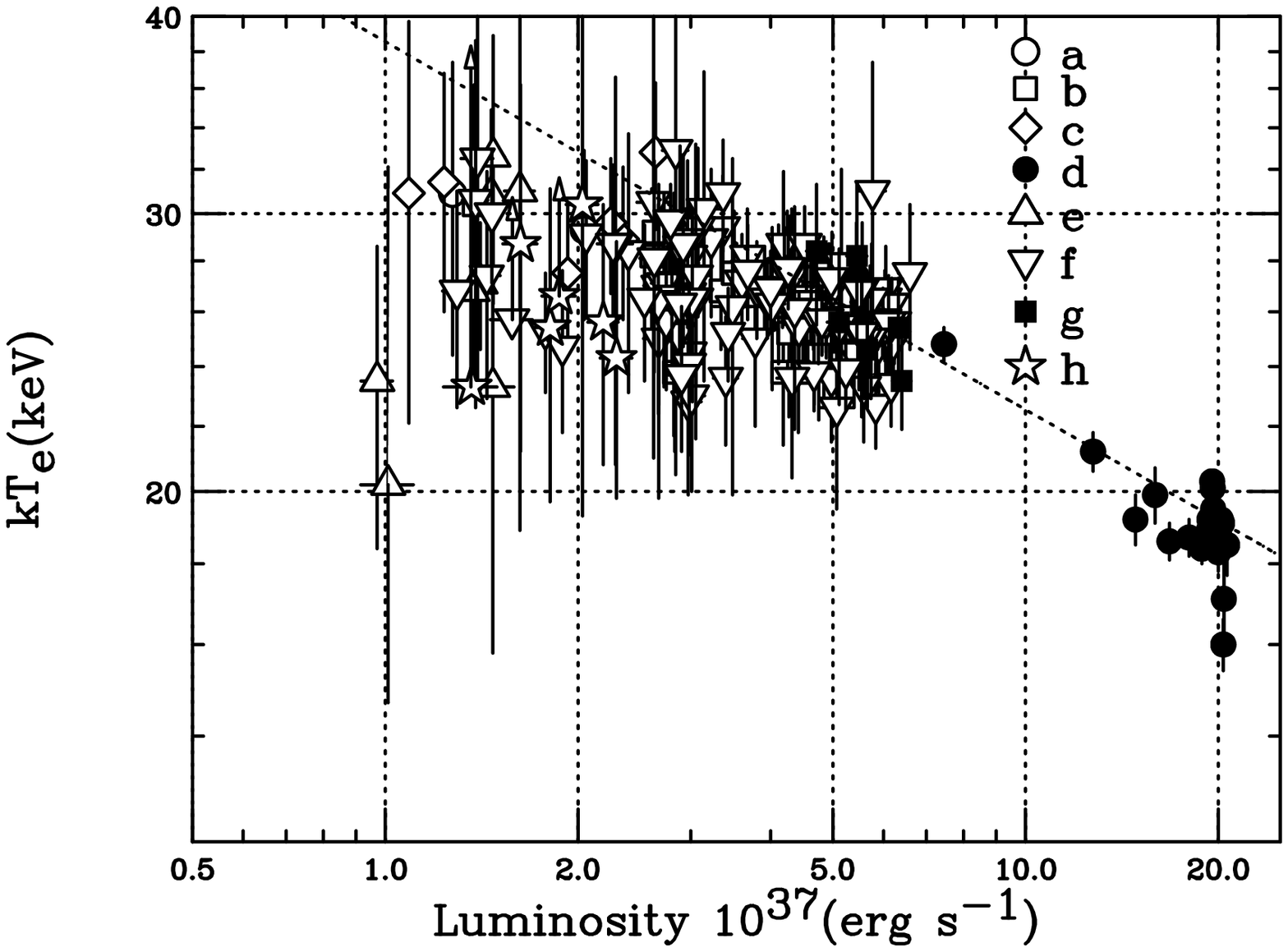} \end{center} \caption{Relation
between luminosity ($L$) in 2-200 keV and energy cut-off
($E_{\rm{cut}}$) obtained by CPL model or electron
temperature($kT_{\rm e}$)  by COMPST model. The dotted lines show the
best-fit relations for $L-E_{cut}$ and $L-kT_e$ using data in the
range of $L>$7$\times$10$^{37}$erg s$^{-1}$.} \label{L_E}
\end{figure}

The top panels of Figures $\ref{L_pho_tau}$ and $\ref{L_E}$  show the
relation between some of the spectral parameters obtained with the CPL
model. The different plot symbols correspond to intervals $a$--$h$ as
defined in Section 2 and as shown in Figure 1.  The high energy
cut-off ($E_{\rm cut}$) ranged from 40 keV to 200 keV or more.
Although we could not constrain the cut-off well at the high energy
end, the highest values we measured are still well below the upper
bound of 500 keV.  As one can see from Figure 6, we found a clear
anti-correlation between  $L$ and $E_{\rm cut}$ when $L > 7 \times
10^{37}$ erg s$^{-1}$.  This relation follows the approximate relation
$E_{\rm cut} \propto L^{-0.70 \pm 0.06}$.  On the other hand, $E_{\rm{cut}}$
seemed to be roughly constant at 200 keV when  $L < 7 \times 10^{37}$
erg s$^{-1}$. To confirm whether this energy cut-off was really 
 required in the HEXTE data of E$\sim$200 keV, we also compared with the
 simple power-law fit results. The F values (=($\Delta \chi^2$/$\Delta
 \nu$)/($\chi^2$/$\nu_2$)) are 23.8, 67.4, and 183 for ($\Delta
 \nu$,$\nu_2$) of the upper three observations shown in Table
 \ref{various_fit}, which indicates that high energy cutoff is required 
 at a confidence level of $>$ 99.9 \%.    
 The periods when the X--ray luminosity was larger than
7$\times 10^{37}$ erg s$^{-1}$ correspond to the initial rising phase
of the outbursts in 2002 (epoch $d$).  Unlike the high energy cut-off, the
photon index was not dependent on the X--ray  luminosity. It was
distributed over 1.4--1.7, which is a typical value for
black hole candidates in the hard state. In the rising phase of the 
 outbursts (epoch $d$ and $f$-$g$), the photon index becomes steeper with
time (1.4 to 1.6), while other epochs show no significant variations
around 1.6--1.7).


In the bottom panels of Figures $\ref{L_pho_tau}$ and $\ref{L_E}$, we
further show the correlations between some of the parameters obtained
from fits with the COMPST model.   The electron temperature $kT_{\rm
e}$ and Thomson optical depth $\tau$ were between $\sim$20--30 keV and
$\sim$4--5, respectively.   We found a clear anti-correlation between
$kT_{\rm e}$ and $L$, $kT_{\rm{e}} \propto L^{-0.24 \pm 0.06}$, when  $L > 7
\times 10^{37}$  erg s$^{-1}$, the same luminosity as was found with
the CPL model. The electron temperature was  more or less constant
around 26 keV  when $L < 7 \times 10^{37}$ erg s$^{-1}$.  No
clear luminosity dependence was found for $\tau$. 

\begin{figure}[t]
\begin{center}
\FigureFile(80mm,50mm){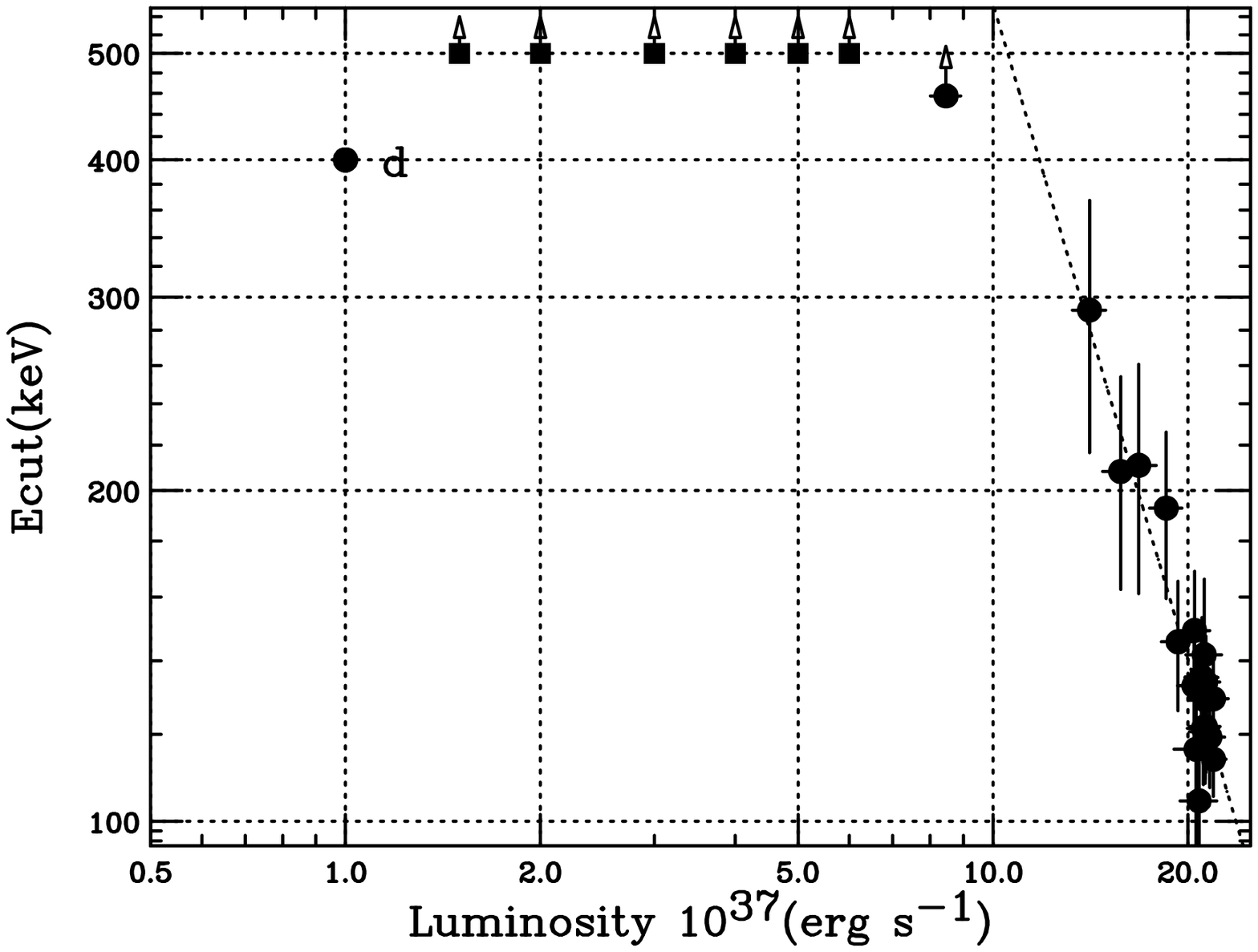}
\FigureFile(80mm,80mm){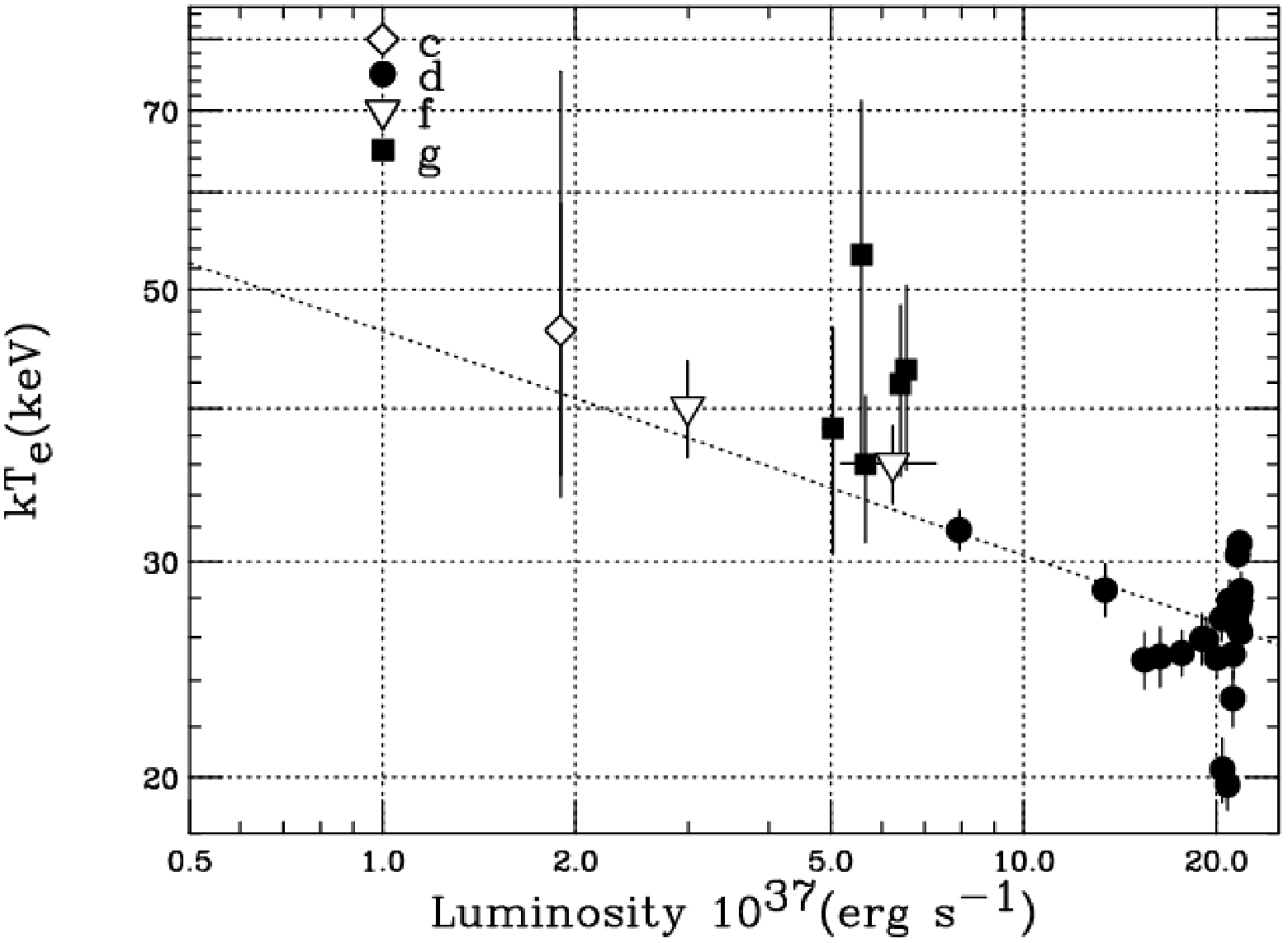}
\end{center}
\caption{Relation between luminosity in 2-200 keV($L$) and high energy
 cutoff/electron temperature after adding a reflection
 component to out cut-off power-law model. The dotted lines show the
 best fit relations for $L-E_{\rm cut}$ and $L-kT_{\rm e}$  using
 the data in the range of $L>$7$\times$10$^{37}$erg s$^{-1}$. }
\label{pexrav_fit}
\end{figure}

So far we have fitted the spectra with relatively simple models. In
other works Compton reflection components have often been taken into
account as well. Therefore, we also checked for a number of our
observations how the fit parameters were affected by using a  model
for reflection from a cold disk (reflect in XSPEC,  Magdziarz $\&$
Zdziarski 1995) plus iron-K emission line with a narrow line width of
50 eV, instead of smeared edge models. The high energy cut-off
($E_{\rm cut}$) and photon index ($\alpha$) became systematically
higher and steeper, respectively, compared to their original fit
values (see Table $\ref{various_fit}$ and top panel of Figure
$\ref{pexrav_fit}$). For example, we derived 1.72$\pm$0.05 (original
value: 1.44$\pm$0.03) and  212$^{+61}_{-40}$  (79.9$^{+7.8}_{-6.7}$)
as the new values of  photon index and $E_{\rm{cut}}$, respectively
for observation ID  70110-01-02-00.  We could not  constrain $E_{\rm
cut}$ in the low luminosity region  ($L<7\times10^{37}$) erg s$^{-1}$
because the reflection component strongly couples with a high energy
cutoff. The same procedure was also applied to the  COMPST model
fitting. The reflection fits gave better results than COMPST with a
smeared edge model in some observations, when $kT_{\rm e}$ was 
lower than 20 keV, although the electron temperature became slightly
higher.  Figure $\ref{pexrav_fit}$ shows the luminosity and
cutoff/electron temperature relation in the epoch $d$ and $g$, which
showed a clear anti-correlation in the non-reflection fits. The
anti-correlation was still found $E_{\rm cut} \propto
L^{-1.98_{-0.60}^{+0.70}}$ or $kT_{\rm{e}} \propto L^{-0.18\pm0.05}$. These 
 correlations have different slopes from other two models, but  
change still seems to occur around the same luminosity, even
when we applied the reflection component,  implying that the change
in behavior above $L=7\times10^{37}$ erg s$^{-1}$ is model
independent.  

The corresponding electron temperature distribution in the
non-reflection fits is in a narrower range, 16--40 keV. These values
should be taken with some care because COMPST model gives a smaller
electron temperature than other Comptonization models, such as the
COMPPS model (Poutanen $\&$ Svensson 1996). We have used the COMPST
model, the simplest and computationally fastest of the various
Comptonization models, but we also tried to apply the COMPPS model
with a spherical geometry  (more complex and accurate) to part of our
data sets for the five observations covering almost all of the
luminosity range from 1.0$\times$10$^{37}$ to 2.1$\times$10$^{37}$
erg s$^{-1}$.  Table \ref{various_fit} shows the  best-fit parameters
for them. The range in electron temperature that we found was
$\sim$40--120 keV, which is consistent with the 50--100 keV range 
found by Zdziarski et al.\ (2004). In addition, the electron
temperature in the bright hard state  ($L>$7$\times$ 10$^{37}$ erg
s$^{-1}$) is significantly lower than that in the low hard state.

\section{Discussion}

\subsection{Summary of Results and Comparison with Previous Results}

We have analyzed a large number of RXTE spectra of the black hole
candidate GX 339--4 when it was in a bright hard state, 
and have found a clear luminosity dependence of the high energy
cut-off in the X-ray spectrum. The 2--200 keV X-ray
luminosities ranged from 1.0 $\times$ 10$^{37}$ to 2.1 $\times$
10$^{38}$ erg s$^{-1}$, covering a factor of 20. This corresponds to
1.3--29\% of Eddington luminosity for a 5.8 solar mass object
(assuming a distance of 8 kpc).  This is the first systematic and
quantitive spectral study over a wide-luminosity range
in the hard state of GX 339--4. We further notice that the
anti-correlation between X-ray luminosity and  energy cut-off was
only found above a luminosity of 7$\times10^{37}$ erg s$^{-1}$.  In
GX 339--4, such luminosities were seen mainly during the rising
phase  of the 2002/2003,  when the source was in a bright hard state.
This state was difficult  to observe prior to RXTE, because typical
black hole transients pass through this state within a few weeks of
the start of an outburst.  RXTE's fast response and flexible
scheduling were critical in observing this state in an increasing 
number of transient BHCs. 

A high energy cut-off around 100 keV has been observed in the energy
spectra of many BHCs in their hard state (Grove et al.\ 1998),
although it is not observed in the hard state of every source as
recent INTEGRAL observations of GRO J1655--40 show (Caballero-Garcia
et al.\ 2007). For GX 339--4 we reconfirmed that it had a high energy
cut-off in  almost all of its hard state observations. It is difficult
to make  comparisons with previous results (Zdziarski et al.\ 1998),
because the authors  used different models, but we clearly found a
varying energy cut-off  in GX 339--4, with values between 40 keV and
200 keV or more.  High energy cut-off energies above 200 keV are
beyond the HEXTE bandpass,  so the highest cut-off energies that we
found may have some intrinsic uncertainties since a part of them 
 were not constrained. Other high  sensitive observations with INTEGRAL or
Suzaku are needed to confirm cut-off energies above 200 keV in GX 339--4. 

\subsection{Anti-correlation between Luminosity and Temperature}

We found that the high energy cut-off or electron temperature
strongly depended on the X-ray luminosity when the luminosity was
higher than 7$\times$10$^{37}$ erg s$^{-1}$, i.e. 0.1 $L_{\rm Edd}$.
A possible anti-correlation in GX 339--4 was already  suggested by
Wardzi$\acute{n}$ski et al.\ (2002) and Zdziarski et al.\ (2004).  A
simple interpretation of the observed anti-correlation, i.e.  that
the hot electrons are efficiently cooled by soft photons via 
inverse-Compton scattering, is  discussed in more detail in the
Section 4.4. Similar anti-correlations have already been suggested 
in GS 2023+238 (Inoue 1994), GRO J0422+32 (Kurfess 1996) and 10 other
BHCs including XTE J1550--564, GS 1354--645, and GX 339--4 (Yamaoka
et al.\ 2005).  Hence, this anti-correlation might be not intrinsic
to GX 339--4  but universal among BHCs.   When the GX 339--4
luminosity was lower than 7$\times$10$^{37}$ erg s$^{-1}$,  the
source seemed to show different behavior; the high energy cut-off
and electron temperature might have a very weak dependence on the
luminosity, remaining a nearly constant at 200 keV and 26 keV,
respectively. The reason why the high  energy cutoff reaches constant
value is beyond the scope of this paper,  but some other cooling
mechanisms such as  bremsstrahlungs and synchrotron emissions may
have to be taken into account.  The hard state of Cygnus X--1, which
has a luminosity of 0.01--0.03 $L_{\rm Edd}$, belongs to the latter
constant-cutoff regime.   This may be the reason why this source has
almost the same cut-off in the hard state at any time (Gierlinski et
al.\ 1997).  


\subsection{Cooling Time Scale due to Inverse Compton Scattering}

Based on the spectral fit results, the thermal Comptonization model
gives a reasonably acceptable fit. In the following, we show whether
the Inverse Compton scattering can be really effective emission 
 process using derived physical parameter.
We consider a spherical hot plasma (like an ADAF for example) with 
a radius of $R$ ($\sim 10
R_{\rm{s}} = 171 (\frac{M}{5.8M_{\odot}})$ km; $R_{\rm{s}}$:
Schwarzchild Radius) around a black hole. In this situation, protons
will be heated up by the gravitational energy release of the
accretion flow through a mechanism such as viscous heating. 
The protons in the plasma lose their energy due to Coulomb collisions
with electrons, while the heated electrons cool due to the Inverse
Compton scattering. 

We define two time scales: $t_{\rm pe}$ ($\propto \frac{ {(\frac{kT_{\rm{p}}}{m_{\rm{p}}} + \frac{k
T_{\rm{e}}}{m_{\rm{e}}})}^{\frac{3}{2}}}{n}$; $n$ is number 
 density of the plasma, $kT_{\rm p}$ is proton temperature and $m_{\rm p}$ is 
 the proton mass.) as 
 the time to establish a Maxwellian velocity distributions for protons
 interacting with electron field particles (for see details Spitzer 1962), and 
$t_{\rm comp}$  ($ = \frac{n k T_{\rm{e}}}{(\frac{dE}{dt})_{\rm comp}}$) as a cooling
time scale of the electron due to inverse Compton scattering. 
The cooling rate per unit volume $(\frac{dE}{dt})_{\rm comp}$ is approximately given as
$\frac{4kT_{\rm{e}}}{m_{\rm{e}} c^2}$ $U_{\rm{rad}} n$ ${\sigma}_{\rm T}$ where
$c$ is the light speed, $U_{\rm{rad}}$ is the photon flux density,
which is given by $U_{\rm rad} \cong \frac{L \tau}{\pi R^2}$,  and
$\sigma_{\rm T}$ is the cross section for Thomson scattering.
These two time scales are given by 
\begin{equation} 
t_{\rm pe} \sim 2.5 \times {10}^{-4} \frac{{(\frac{k T_{\rm{e}}}{30
 keV})}^{\frac{3}{2}} (\frac{R}{10R_{\rm{s}}})}{{(\frac{\tau}{5})}} [{\rm sec}],
 \end{equation}
 and
\begin{equation} 
t_{\rm comp} \sim 5.7 \times {10}^{-6}
 \frac{{(\frac{R}{10R_{\rm{s}}})}^2}{(\frac{L}{{10}^{37} erg s^{-1}})(\frac{\tau}{5})} [{\rm sec}],
\end{equation}
respectively. On the other hand, $t_{\rm vis}$($ \sim \frac{R^2}{\alpha H^2
\Omega}$), the viscous time scale in the ADAF, that is, the time scale in which 
electrons fall to the central black hole, is
\begin{equation}
t_{\rm vis} \sim 2.6 \times {10}^{-3}\frac{{(\frac{R}{H})}^2 {(\frac{R}{10 R_{\rm{s}}})}^{3/2}}{\alpha} [{\rm sec}],
\end{equation}
where 2$H$ is the thickness of disk, $\Omega$ is the angular velocity 
(= $\sqrt{\frac{GM}{R^3}}$), and where we
assumed $\frac{R}{H}\sim 1$ and $\alpha \sim 1$ ($\alpha$ is viscous
parameter) for the hard state.  Clearly, for reasonable parameters,
$t_{\rm pe}$ and $t_{\rm comp}$ are much smaller than $t_{\rm vis}$, which 
means that inverse Compton scattering is an efficient cooling process 
in this situation.


\subsection{Qualitative Explanation for Anti-correlation between Luminosity and Electron Temperature}

Next, we try to explain the anti-correlation between $L$ and 
$kT_{\rm e}$ quantitavely in the following simple discussion. 
The proton temperature ($kT_{\rm p}$) is assumed to be 
approximately constant at $\sim \frac{GM m_{\rm p}}{R}$, 
 that is, the energy loss rate of protons is 
much smaller than viscous heating rate through the accretion.  
These protons will give their energy to electrons through collisions and 
the energy loss rate per unit volume is given as $\frac{\frac{3}{2}nkT_{\rm
p}}{t_{\rm{pe}}}$ .
 If $\frac{k T_{\rm{e}}}{m_e c^2}$ is larger than $\frac{k T_p}{m_p
 c^2}$,  we will get 

\begin{equation}
t_{\rm pe} \propto \frac{T_{\rm{e}}^{\frac{3}{2}}}{n}
\end{equation}

 In a steady state, the heating
rate from proton to electron  should balance with the cooling rate due to
inverse Compton scattering, i.e.  
\begin{equation}
\frac{\frac{3}{2} nk T_{\rm{p}}}{t_{\rm{pe}}} = \frac{4k T_{\rm{e}}}{m_{\rm{e}} c^2} U_{\rm{rad}} n {\sigma}_{\rm T}
\end{equation}
Using these equation (4) and (5), we can work out the anti-correlation of
$k T_{\rm{e}} \propto L^{-\frac{2}{5}}$ quantitatively. This is
attributed to the fact that the radiation mechanism in the hard
state is due to inverse Compton scattering. 

Zdziarski (1998) suggested that the high energy cut-off depends on the
luminosity from a theoretical point of view of thermal Comptonization
by a more precise model. The relations  $k T_{\rm{e}} \propto L^{-2/7}$
and $k T_{\rm{e}} \propto L^{-1/6}$ were derived for the advection
dominated and cooling dominated cases, respectively.  In comparison
with this prediction, our result, $k T_{\rm{e}} \propto L^{-0.24\pm0.06}$,
is close to  these values. Furthermore, the maximum luminosity in
the low/hard state  is predicted as $L_{\rm max} \sim 0.15
y^{\frac{3}{5}} \alpha^{\frac{7}{5}} L_{\rm Edd}$ in the model of
Zdziarski (1998). The highest luminosity observed in 2002,
$\sim$0.29$L_{\rm Edd}$, agrees within a factor of 2 with
this value.

In recent years the X-ray spectra of the hard state have also been
discussed in the framework of jet outflows. In the most recent version
of this model (Markoff, Nowak, \& Wilms 2005) the hard X-ray spectra
are dominated by synchrotron (at the low end) and synchrotron
self-Compton (SSC) emission (at the high end) from the magnetized
plasma at the base of a jet, which shares many properties with other
proposed types of Comptonizing coronae.  In the simplest jet SSC
interpretation, the observed decrease of the energy cut-off with
luminosity means that either the size of the base of the jet is
increasing or that the electron temperature is decreasing as the
luminosity goes up. Changes are also expected in the synchrotron
cut-off, which should decrease with luminosity, because of a higher
cooling rate. There is likely an interplay between the synchrotron and
the SSC components that needs to be worked out in more detail before
the jet model can be applied to our data.

\subsection{Difference between bright hard state and common low/hard state}

Differences between the bright hard state (i.e.\ $\sim$0.2 $L_{Edd}$)
and the more common low/hard state have not been widely discussed in
the literature.  From an observational point of view, we found that
the spectral properties in the bright hard state were not very
different from the typical low/hard state except for one point: the
high energy cut-off in the bright hard state is considerable lower
($E_{cut}\approx$50 keV) than in the low hard state
($E_{cut}\approx$200 keV). The timing properties are also suggested to
be almost the same, with a moderate increase of the frequencies with
luminosity (Belloni et al. 2005). A clear correlation between the radio
and X-ray flux has been established  in the hard state of GX 339--4 (Corbel
et al.\ 2003), but this relation has only been measured up to a
luminosity of $\sim6\times10^{37}$ erg s$^{-1}$, i.e. just up to the
luminosity above which we see a decrease in the energy cut-off. 

The low/hard state has been considered to correspond to an
accretion-disk solution of the optically thin ADAF (Esin et al.\ 1997).
Observed luminosity ($\sim$ 0.2L$_{\rm Edd}$) in the bright hard state can
only be explained by this solution (0.4$\alpha^2 L_{\rm Edd}$) if we assume a large viscosity
parameter $\alpha \sim 1$. 
It is difficult to estimate an actual value of $\alpha$, but recent three-dimensional 
 MHD simulations suggest that $\alpha$ takes a value of reasonably 10$^{-2}$ to
 10$^{-1}$ (Hawley \& Krolik 2001).  Furthermore, considering the presence 
of hysteresis in the state transition, i.e. the
fact that the bright hard state is only in the rise phase of the
outburst, the bright hard state may  suggest a presence of another
solution like luminous hot accretion flow (LHAF: Yuan
2001) rather than an ADAF.  Further investigations including the
radio/X-ray correlations are needed for the bright hard state through
multi-wavelength observations.  

\subsection{Constancy of the Photon Index Variations}

\begin{figure}[htbp]
\begin{center}
\FigureFile(80mm,80mm){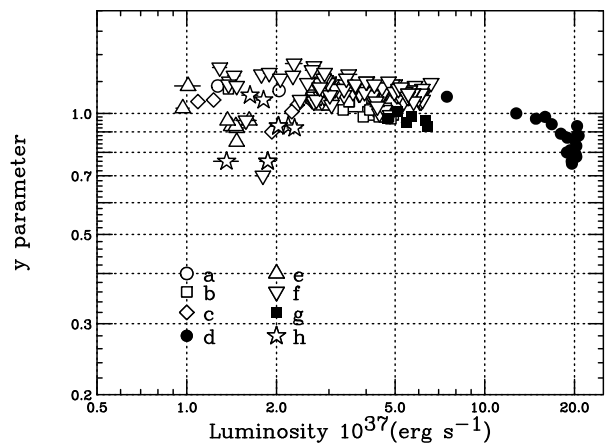}
\end{center}
\caption{Relation between the luminosity($L$) and $y$ parameter obtained
 by COMPST model. The different symbols correspond to the epoch defined
 in Figure \ref{lc_1996_2005}. }
\label{L_y}
\end{figure}

Another important result of our systematic study is that the slope of 
photon index remains almost constant at 1.4--1.7 regardless of the
source intensity. This value is typical for the hard state of
black hole candidates.  From the point of view of thermal
Comptonization, the photon index is correlated with Compton $y$
parameter (= $\frac{kT_e}{m_{\rm{e}} c^2}$max$(\tau,\tau^2)$; Sunyaev
\& Titartuck 1980). Thus, the obtained $y$ parameter distribution is
remarkably close to unity for more than an order of  magnitude as seen in
Figure \ref{L_y}. This is consistent with  power-law shape of the
spectrum in the hard state. In the thermal  Comptonization model,
the $y$ parameter should always be close to be  unity when there is a
large number of soft photons (Shapiro, Lightman,  and Eardly 1976).
The constancy of the $y$ parameter means that the ratio of
Comptonization luminosity to that in the seed soft photons  is
constant, indicating a constant geometry of the corona.

\section{Summary}

We have analyzed 200 RXTE observations of the black hole
candidate GX 339--4 in the hard state during its outbursts in 1996-2005.
The observed luminosities ranged was from 1.0$\times$10$^{37}$ erg
s$^{-1}$ to 2.1$\times$10$^{38}$ erg s$^{-1}$, assuming a source
distance at 8 kpc. All the broadband spectra are well-explained by
both cut-off power law (CPL) model and thermal Comptonization
(COMPST) model with a spherical geometry.   The photon index (in the
cut-off power-law model)  ranged from 1.4--1.7 regardless of the
luminosity. The Compton $y$ parameter remained also close to unity
regardless of the intensity variations, which is consistent with its
power-law like spectral shape. The high energy cut-off
in the CPL model ranges from 40 to 280 keV or more, and electron
temperature in the COMPST model from 16 to 40 keV.  Both parameters  
depend on the luminosity. We found a clear anti-correlation of 
$E_{\rm{cut}}$ $\propto$ $L^{-0.70 \pm 0.06}$ between $L$ and 
$E_{\rm{cut}}$ or $kT_{\rm{e}}$ $\propto$ $L^{-0.24 \pm 0.06}$ between $L$
and $kT_{\rm{e}}$ when the $L$ is 7$\times$10$^{37}$ erg s$^{-1}$. This
former anti-correlation is quantitatively explained by the scenario
that the  heating rate from protons to electrons balances with
cooling rate of  inverse Compton scattering. The cooling time scale
of inverse Compton  scattering obtained by spectral parameter is much
smaller than  viscous time scale, suggesting that the inverse Compton
scattering is dominant in the radiation mechanisms.  
Further detailed studies of a high energy cut-off
using INTEGRAL and Suzaku mission or future high energy missions 
will be valuable and interesting.
\\

The authors would like to thank the referee, Dr. Aya Kubota, for her valuable
suggestions and comments. They also thank Sera Markoff for discussions on our results with
regard to jet models for the hard state. 
This research has made use of data obtained through the
High Energy Astrophysics Science Archive Research Center Online
Service, provided by the NASA/Goddard Space Flight Center.

\setlength{\tabcolsep}{2.4pt}
\begin{table*}
\begin{center}
\rotatebox{90}{
\begin{minipage}[t]{\textwidth}
\scriptsize
\caption{Results of spectral fitting (I).}
\begin{tabular}{c c c c c c c c c c c c}
\hline
\hline
 & & cut-off power-law & (CPL) & & thermal & Comptonization & (COMPST) & & \\
\hline
Observation ID & Observation Time & Flux (erg/sec/cm$^2$)  & $E_{\rm{cut}}$ (keV) &
 photon index & MAX$\tau$ & ${\chi}^2$/d.o.f & Flux (erg/sec/cm$^2$) &
 k$T_{\rm{e}}$ (keV) & $\tau$ & MAX$\tau$ & ${\chi}^2$/d.o.f  \\
\hline
10420-01-01-00 & 1996-07-26 18:20-20:15 & (6.5$\pm$0.3)$\times 10^{-9}$
 & $184.2_{-28.6}^{+40.0}$ & 1.52$\pm$0.03 & 1.45$\pm$0.24 & 80.8/79 &
 $6.2_{-0.1}^{+0.2} \times 10^{-9}$ & $26.00_{-1.36}^{+1.56}$ & 4.44$\pm$0.17 & 0.71$\pm$0.19 & 90.1/79 \\
10068-05-01-00 & 1996-10-17 02:12-05:07 & (1.7$\pm$0.1)$\times 10^{-9}$
 & -- & $1.49_{-0.05}^{+0.04}$ & $0.96_{-0.36}^{+0.39}$ & 74.9/79 & (1.7$\pm$0.1)$\times 10^{-9}$ & $30.88_{-4.76}^{+8.21}$ & $4.40_{-0.58}^{+0.47}$ & 0.60$\pm$0.33 & 77.4/79 \\
10068-05-02-00 & 1996-10-29 22:34-00:45 & $2.8_{-0.2}^{+0.1} \times 10^{-9}$ & $>$251.9 & $1.50_{-0.04}^{+0.03}$ & $0.96_{-0.31}^{+0.34}$
 & 64.3/79 & (2.7$\pm$0.1)$\times 10^{-9}$ & $29.34_{-3.02}^{+4.14}$ & $4.46_{-0.36}^{+0.33}$ & $0.55_{-0.28}^{+0.28}$ & 61.6/79 \\
20181-01-01-01 & 1997-02-03 15:56-19:09 & (6.6$\pm$0.3)$\times 10^{-9}$
 & $207.0_{-32.5}^{+46.0}$ & 1.52$\pm$0.02 & 1.41$\pm$0.23 & 79.2/79 &
 (6.3$\pm$0.1)$\times 10^{-9}$ & $27.52_{-1.39}^{+1.58}$ & 4.3$\pm$0.16 & 0.65$\pm$0.18 & 91.5/79 \\
20181-01-01-00  & 1997-02-03 22:27-01:36 & (6.4$\pm$0.3)$\times 10^{-9}$ & $193.8_{-31.1}^{+44.5}$ & 1.52$\pm$0.03 & 1.46$\pm$0.25 & 62.0/79 & (6.1$\pm$0.2)$\times 10^{-9}$ & $26.70_{-1.42}^{+1.63}$ & $4.38_{-0.18}^{+ 0.17}$ & 0.69$\pm$0.20 & 70.7/79 \\ 
20183-01-01-00 & 1997-02-08 14:20-20:48 & (6.1$\pm$0.2)$\times 10^{-9}$
 & $167.7_{-18.8}^{+24.2}$ & 1.50$\pm$0.02 & 1.71$\pm$0.21 & 82.7/79 &
 (5.7$\pm$0.1)$\times 10^{-9}$	& $26.19_{-0.99}^{+1.09}$ & 4.43$\pm$0.13 & 0.75$\pm$0.17 & 79.4/79 \\
20181-01-02-00 & 1997-02-10 15:49-20:51 & (5.9$\pm$0.2)$\times 10^{-9}$
 & $178.2_{-23.5}^{+31.1}$ & 1.49$\pm$0.02 & 1.61$\pm$0.22 & 63.1/79 &
 (5.5$\pm$0.1)$\times 10^{-9}$ & $26.20_{-1.14}^{+1.27}$ & 4.48$\pm$0.15 & 0.73$\pm$0.18 & 73.3/79 \\
20183-01-02-00 & 1997-02-14 00:09-06:43 & (5.9$\pm$0.3)$\times 10^{-9}$
 & $243.4_{-49.4}^{+81.0}$ & 1.52$\pm$0.03 & 1.48$\pm$0.26 & 95.9/79 &
 (5.5$\pm$0.1)$\times 10^{-9}$	& $28.16_{-1.73}^{+2.05}$ & $4.29_{-0.20}^{+0.19}$ & 0.72$\pm$0.20 & 92.8/79 \\
20183-01-02-01 & 1997-02-14 14:20-21:22 & (5.9$\pm$0.2)$\times 10^{-9}$
 & $168.6_{-18.4}^{+23.0}$ & 1.47$\pm$0.02 & 1.85$\pm$0.21 & 115.6/79 &
 (5.5$\pm$0.1)$\times 10^{-9}$	& $26.44_{-0.95}^{+1.03}$ & 4.46$\pm$0.12 & 0.79$\pm$0.16 & 111.1/79 \\
20181-01-03-00 & 1997-02-17 18:29-00:12 & (5.8$\pm$0.3)$\times 10^{-9}$
 & $162.5_{-25.5}^{+35.8}$ & 1.47$\pm$0.04 & $1.80_{-0.31}^{+0.32}$ &
 77.4/79 & (5.3$\pm$0.2)$\times 10^{-9}$ & $26.94_{-1.48}^{+1.71}$ & 4.37$\pm$0.19 & 0.60$\pm$0.23 & 80.8/79 \\
20056-01-01-00 & 1997-04-05 08:36-09:15 & (5.6$\pm$0.3)$\times 10^{-9}$ & $237.3_{-64.4}^{+132.8}$ & 1.54$\pm$0.04 & 1.20$\pm$0.30 & 64.8/79 & (5.4$\pm$0.2)$\times 10^{-9}$ & $26.84_{-2.38}^{+3.03}$ & $4.41_{-0.29}^{+0.27}$ & $0.73_{-0.25}^{+0.25}$ & 68.2/79 \\
20056-01-02-00 & 1997-04-10 11:47-12:28 & (6.2$\pm$0.3)$\times 10^{-9}$ & $179.6_{-37.4}^{+60.5}$
 & 1.55$\pm$0.03 & $1.30_{-0.29}^{+0.28}$ & 68.2/79 & (6.0$\pm$0.2)$\times 10^{-9}$ & $25.71_{-1.97}^{+2.39}$ & $4.42_{-0.25}^{+0.24}$ & 0.72$\pm$0.24 & 77.9/79 \\
20056-01-03-00 & 1997-04-11 13:26-14:06 & (6.3$\pm$0.3)$\times 10^{-9}$ & $165.7_{-31.2}^{+47.5}$
 & 1.51$\pm$0.03 & 1.43$\pm$0.28 & 79.0/79 & (6.1$\pm$0.2)$\times 10^{-9}$ & $24.45_{-1.66}^{+1.94}$ & 4.64$\pm$0.23 & 0.83$\pm$0.24 & 83.1/79 \\
20056-01-06-00 & 1997-04-17 23:26-00:01  & (6.5$\pm$0.3)$\times 10^{-9}$ & $167.5_{-34.6}^{+55.0}$ & 1.55$\pm$0.03 & 1.43$\pm$0.28 & 56.1/79 & (6.4$\pm$0.2)$\times 10^{-9}$ & $25.93_{-2.42}^{+3.09}$ & $4.42_{-0.30}^{+0.28}$ & 1.03$\pm$0.24 & 55.6/79 \\
20056-01-06-01 & 1997-04-18 00:01-00:09 & $7.4_{-0.6}^{+0.5}$ $\times
 10^{-9}$ & $>$ 198.4 & $1.59_{-0.06}^{+0.04}$ & $1.29_{-0.42}^{+0.45}$ &
 68.2/79 & (6.3$\pm$0.2)$\times 10^{-9}$ & $24.89_{-2.03}^{+2.48}$ & $4.46_{-0.27}^{+0.26}$ & 0.84$\pm$0.24 & 73.3/79 \\
20056-01-07-00 & 1997-04-19 22:20-23:21 & (6.5$\pm$0.3)$\times 10^{-9}$ & $170.9_{-36.4}^{+60.1}$ &
 1.55$\pm$0.03 & $1.54_{-0.29}^{+0.28}$ & 52.77/79 & (6.2$\pm$0.2)$\times 10^{-9}$ & $24.41_{-1.98}^{+2.40}$ & $4.53_{-0.27}^{+0.26}$ & 1.00$\pm$0.24 & 60.7/79 \\
20056-01-08-00 & 1997-04-22 21:53-22:29 & (6.0$\pm$0.3)$\times 10^{-9}$ & $162.0_{-33.2}^{+52.6}$ & 1.56$\pm$0.04 & 1.54$\pm$0.29 & 71.6/79 & (5.8$\pm$0.2)$\times 10^{-9}$ & $23.81_{-1.82}^{+2.18}$ & $4.58_{-0.26}^{+0.25}$ & $1.02_{-0.24}^{+0.25}$ & 73.5/79 \\
20181-01-04-00 & 1997-05-29 09:23-13:51 & (3.9$\pm$0.2)$\times 10^{-9}$
 & $203.9_{-39.3}^{+60.9}$ & 1.49$\pm$0.03 & 1.41$\pm$0.29 & 78.1/79 &
 (3.6$\pm$0.1)$\times 10^{-9}$ & $27.10_{-1.74}^{+2.08}$ & $4.44_{-0.22}^{+0.21}$ & $0.60_{-0.22}^{+0.23}$ & 92.1/79 \\
20181-01-05-00 & 1997-07-07 08:16-13:01 & (1.9$\pm$0.1)$\times 10^{-9}$
 & $>$232.7 & $1.49_{-0.04}^{+0.03}$ & $1.00_{-0.31}^{+0.34}$ & 65.8/79
 & (1.8$\pm$0.1)$\times 10^{-9}$ & $30.43_{-4.24}^{+7.33}$ & $4.40_{-0.53}^{+0.42}$ & $0.55_{-0.28}^{+0.29}$ & 80.5/79 \\
20181-01-06-00 & 1997-08-23 04:23-07:29 & (4.9$\pm$0.3)$\times 10^{-9}$
 & $235.3_{-61.4}^{+120.4}$ & 1.50$\pm$0.04 & 1.62$\pm$0.35 & 72.5/79 &
 (4.6$\pm$0.2)$\times 10^{-9}$ & $27.48_{-2.31}^{+2.95}$ & $4.45_{-0.30}^{+0.28}$ & 0.91$\pm$0.28 & 76.2/79 \\
20181-01-06-01 & 1997-08-23 11:04-14:46 & (4.8$\pm$0.2)$\times 10^{-9}$ & $241.2_{-51.8}^{+87.5}$ & 1.50$\pm$0.03 & 1.37$\pm$0.27 & 53.0/79 & (4.6$\pm$0.1)$\times 10^{-9}$ & $27.48_{-1.77}^{+2.10}$ & $4.46_{-0.22}^{+0.21}$ & 0.73$\pm$0.21 & 55.2/79 \\
20181-01-07-01 & 1997-09-19 22:09-22:37 & (5.7$\pm$0.3)$\times 10^{-9}$
 & $179.8_{-45.9}^{+86.2}$ & 1.52$\pm$0.04 & 1.37$\pm$0.32 & 46.7/79 &
 (5.5$\pm$0.2)$\times 10^{-9}$ 	& $23.79_{-2.11}^{+2.59}$ & $4.79_{-0.31}^{+0.30}$ & $0.94_{-0.27}^{+0.28}$ & 42.0/79 \\
20181-01-07-02 & 1997-09-19 23:57-00:32 & (5.8$\pm$0.4)$\times 10^{-9}$
 & $237.3_{-77.4}^{+195.7}$ & 1.54$\pm$0.04 & $1.36_{-0.34}^{+0.35}$ &
 58.4/79 & (5.6$\pm$0.2)$\times 10^{-9}$ & $25.19_{-2.53}^{+3.25}$ & $4.63_{-0.35}^{+0.33}$ & 1.06$\pm$0.30 & 47.3/79 \\
20181-01-07-00 & 1997-09-20 01:30-05:58 & (5.8$\pm$0.3)$\times 10^{-9}$
 & $176.2_{-27.6}^{+39.0}$ & 1.50$\pm$0.03 & 1.87$\pm$0.28 & 77.5/79 &
 (5.4$\pm$0.1)$\times 10^{-9}$ & $27.27_{-1.48}^{+1.71}$ & 4.31$\pm$0.18 & 0.86$\pm$0.21 & 82.3/79 \\
20183-01-03-00 & 1997-10-22 03:01-05:52 & (5.4$\pm$0.2)$\times 10^{-9}$
 & $213.3_{-39.3}^{+60.4}$ & 1.55$\pm$0.03 & 1.43$\pm$0.24 & 58.6/79 &
 (5.1$\pm$0.1)$\times 10^{-9}$ 	& $26.93_{-1.60}^{+1.88}$ & $4.32_{-0.19}^{+0.18}$ & 0.80$\pm$0.20 & 62.8/79 \\
20183-01-04-00 & 1997-10-25 03:22-06:16 & $4.6_{-0.2}^{+0.3} \times
 10^{-9}$  & $204.3_{-43.2}^{+71.1}$ & 1.52$\pm$0.04 & 1.39$\pm$0.30 &
 60.8/79 & (4.4$\pm$0.1)$\times 10^{-9}$ & $26.78_{-1.95}^{+2.35}$ & $4.41_{-0.25}^{+0.24}$ & 0.70$\pm$0.24 & 71.2/79 \\
20181-01-08-000 & 1997-10-28 03:00-09:46 & (4.1$\pm$0.2)$\times 10^{-9}$ & $217.4_{-40.7}^{+63.1}$ & 1.50$\pm$0.03 & 1.39$\pm$0.27 & 81.3/79 & (3.9$\pm$0.1)$\times 10^{-9}$ & $27.52_{-1.58}^{+1.85}$ & $4.41_{-0.20}^{+0.19}$ & 0.60$\pm$0.21 & 74.8/79 \\
20181-01-08-00 & 1997-10-28 09:46-11:46 & $3.9_{-0.2}^{+0.3} \times
 10^{-9}$  & $212.6_{-64.8}^{+151.2}$ & 1.51$\pm$0.04 & 1.11$\pm$0.35 &
 74.1/79 & (3.7$\pm$0.1)$\times 10^{-9}$ & $23.79_{-2.31}^{+2.92}$ & $4.88_{-0.35}^{+0.33}$ & 0.73$\pm$0.30 & 63.4/79 \\
20183-01-05-00 & 1997-10-28 18:08-22:13 & (4.0$\pm$0.2)$\times 10^{-9}$
 & $215.2_{-48.4}^{+85.9}$ & 1.48$\pm$0.04 & 1.43$\pm$0.31 & 73.3/79 &
 (3.8$\pm$0.1)$\times 10^{-9}$ & $25.60_{-1.75}^{+2.06}$ & 4.70$\pm$0.24 & 0.81$\pm$0.25 & 58.0/79 \\
20183-01-06-00 & 1997-10-31 20:35-23:48 & (3.6$\pm$0.2)$\times 10^{-9}$
 & $>$210.7 & 1.52$\pm$0.04 & 1.06$\pm$0.30 & 49.8/79 &
 (3.4$\pm$0.1)$\times 10^{-9}$ & $28.23_{-2.52}^{+3.21}$ & $4.44_{-0.29}^{+0.28}$ & 0.62$\pm$0.25 & 45.5/79 \\
20183-01-07-00 & 1997-11-03 20:35-23:48 & (3.6$\pm$0.2)$\times 10^{-9}$
 & $>$254.3 & 1.51$\pm$0.03 & 1.12$\pm$0.27 & 56.1/79 &
 (3.4$\pm$0.1)$\times 10^{-9}$ & $29.16_{-2.30}^{+2.93}$ & $4.41_{-0.26}^{+0.24}$ & 0.65$\pm$0.23 & 51.9/79 \\
40108-01-03-00 & 1999-02-12 21:40-01:54 & (2.6$\pm$0.2)$\times 10^{-9}$
 & $249.9_{-69.7}^{+144.0}$ & 1.62$\pm$0.04 & 1.56$\pm$0.29 & 60.2/79 &
 (2.5$\pm$0.1)$\times 10^{-9}$	& $27.54_{-2.62}^{+3.43}$ & $4.09_{-0.31}^{+0.29}$ & 1.13$\pm$0.27 & 52.7/79 \\
40108-01-04-00 & 1999-03-03 21:28-00:39 & (3.2$\pm$0.2)$\times 10^{-9}$
 & $>$225.1 & $1.54_{-0.04}^{+0.02}$ & 1.43$\pm$0.32 & 64.1/79 & (3.1$\pm$0.1)$\times 10^{-9}$	& $28.93_{-2.77}^{+3.65}$ & $4.31_{-0.31}^{+0.29}$ & $0.96_{-0.25}^{+0.26}$ & 58.0/79 \\
40108-01-04-000 & 1999-03-03 15:07-21:28 & (3.2$\pm$0.2)$\times 10^{-9}$ & $>$223.0 & 1.54$\pm$0.04 & 1.23$\pm$0.33 & 62.7/79 & (2.9$\pm$0.1)$\times 10^{-9}$ & $29.55_{-2.53}^{+3.28}$ & $4.17_{-0.28}^{+ 0.26}$ & 0.60$\pm$0.25 & 59.9/79 \\  
40108-02-01-00 & 1999-04-02 13:05-17:46 & (3.7$\pm$0.2)$\times 10^{-9}$
 & $>$254.9 & $1.49_{-0.04}^{+0.03}$ & 1.52$\pm$0.33 & 52.0/73 & (3.4$\pm$0.1)$\times 10^{-9}$	& $32.82_{-2.98}^{+4.11}$ & $4.09_{-0.30}^{+0.27}$ & 0.77$\pm$0.25 & 66.9/73 \\
40108-02-02-00 & 1999-04-22 10:46-17:09 & (1.7$\pm$0.1)$\times 10^{-9}$
 & $>$226.8 & $1.51_{-0.05}^{+0.04}$ & $1.13_{-0.20}^{+0.43}$ & 61.0/73
 & (1.6$\pm$0.1)$\times 10^{-9}$ & $31.41_{-4.11}^{+6.61}$ & $4.20_{-0.48}^{+0.41}$ & $0.56_{-0.33}^{+0.34}$ & 66.0/73 \\
40104-01-01-00 & 1999-04-27 04:18-08:41 & (1.4$\pm$0.1)$\times 10^{-9}$
 & $>$174.6 & 1.52$\pm$0.05 & $0.93_{-0.40}^{+0.41}$ & 57.7/73 & (1.4$\pm$0.1)$\times 10^{-9}$	& $30.91_{-5.42}^{+12.20}$ & $4.21_{-0.76}^{+0.53}$ & $0.45_{-0.34}^{+0.35}$ & 69.8/73 \\
60705-01-55-00 & 2002-04-02 14:24-14:51 & (7.5$\pm$0.5)$\times 10^{-9}$
 & $203.9_{-45.0}^{+76.0}$ & 1.44$\pm$0.04 & 1.28$\pm$0.35 & 80.8/73 &
 $7.1_{-0.2}^{+0.3} \times 10^{-9}$ & $27.43_{-2.01}^{+2.45}$ & $4.61_{-0.27}^{+0.26}$ & $0.41_{-0.27}^{+0.28}$ & 93.7/73 \\
70109-01-02-00 & 2002-04-03 18:16-21:37 & $1.1_{-0.03}^{+0.04} \times
 10^{-8}$ & $114.9_{-6.9}^{+7.8}$ & 1.36$\pm$0.02 &
 $1.94_{-0.19}^{+0.19}$ & 79.9/73 & (9.7$\pm$0.2)$\times 10^{-9}$ & $24.83_{-0.57}^{+0.59}$ & $4.76_{-0.09}^{+0.09}$ & $0.26_{-0.15}^{+0.15}$ & 140.9/73 \\
70109-01-01-00 & 2002-04-07 13:07-14:07 & (1.8$\pm$0.1)$\times 10^{-8}$
 & $84.9_{-5.6}^{+6.3}$ & 1.41$\pm$0.02 &
 $1.93_{-0.23}^{+0.23}$ & 66.5/73 & (1.7$\pm$0.04)$\times 10^{-8}$ & $21.25_{-0.63}^{+0.67}$ & $4.90_{-0.12}^{+0.12}$ & $0.46_{-0.19}^{+0.19}$ & 139.5/73 \\
70110-01-01-00 & 2002-04-09 05:42-06:21 & (2.0$\pm$0.1)$\times 10^{-8}$
 & $75.4_{-5.8}^{+6.6}$ & 1.44$\pm$0.03 & 1.80$\pm$0.25 & 62.9/73 & (1.9$\pm$0.1)
 $\times 10^{-8}$ & $19.21_{-0.71}^{+0.76}$ & $5.09_{-0.15}^{+0.16}$ & 0.54$\pm$0.21 & 111.5/73 \\
70110-01-02-00 & 2002-04-10 10:11-10:22 & (2.2$\pm$0.1)$\times 10^{-8}$
 & $79.9_{-6.7}^{+7.8}$ & 1.44$\pm$0.03 & 1.79$\pm$0.27 & 81.5/73 &
 (2.1$\pm$0.1)$\times 10^{-8}$ & $19.95_{-0.78}^{+0.85}$ & 5.02$\pm$0.16 & $0.60_{-0.22}^{+0.23}$ & 125.7/73 \\
\hline
\multicolumn{4}{@{Errors are 90$\%$} @{confidence.}  @{The unabsorbed
 flux}  @{is calculated for 2-200 keV.}}{\hbox to 0pt {\parbox{85mm}}{\footnotesize
  \footnotemark[]
}\hss}
\label{table_1}
\end{tabular}
\end{minipage}}
\end{center}
\end{table*}


\setlength{\tabcolsep}{2.4pt}
\begin{table*}
\rotatebox{90}{
\begin{minipage}[t]{\textwidth}
\scriptsize
\begin{center}
\caption{Results of spectral fitting (II).}
\begin{tabular}{c c c c c c c c c c c c}
\hline
\hline
 & & cut-off power-law & (CPL) & & & & thermal & Comptonization & (COMPST) & & \\  
\hline
Observation ID & Observation Time & Flux (erg/sec/cm$^2$) & $E_{\rm{cut}}$ (keV) &
 photon index & MAX$\tau$ & ${\chi}^2$/d.o.f & Flux (erg/sec/cm$^2$) &
 k$T_{\rm{e}}$ (keV) & $\tau$ & MAX$\tau$ & ${\chi}^2$/d.o.f  \\\hline
70109-01-03-00 & 2002-04-13 02:06-02:30 & (2.3$\pm$0.1)$\times 10^{-8}$
 & $65.7_{-3.5}^{+3.8}$ & $1.41_{-0.02}^{+0.02}$ &
 $2.10_{-0.21}^{+0.21}$ & 58.4/73 & (2.2$\pm$0.1)$\times 10^{-8}$ & $18.62_{-0.49}^{+0.52}$ & $5.08_{-0.11}^{+0.11}$ & $0.52_{-0.18}^{+0.18}$ & 165.5/73 \\
70110-01-03-00 & 2002-04-15 03:13-03:40 & (2.5$\pm$0.1)$\times 10^{-8}$
 & $63.3_{-3.1}^{+3.4}$ & 1.44$\pm$0.02 & 1.83$\pm$0.21 & 46.8/73 &
 (2.4$\pm$0.1)$\times 10^{-8}$ & $18.74_{-0.48}^{+0.50}$ & 4.93$\pm$0.11 & $<$0.32 & 241.2/73 \\
70110-01-04-00 & 2002-04-17 02:43-03:09 & (2.6$\pm$0.1)$\times 10^{-8}$
 & $60.9_{-2.7}^{+2.9}$ & 1.45$\pm$0.02 & 2.06$\pm$0.20 & 80.3/73 &
 (2.5$\pm$0.1)$\times 10^{-8}$ & $18.47_{-0.42}^{+0.44}$ & 4.91$\pm$0.10 & 0.33$\pm$0.17 & 260.7/73 \\
40031-03-01-00 & 2002-04-18 02:28-03:12 & (2.6$\pm$0.1)$\times 10^{-8}$
 & $61.6_{-2.7}^{+2.9}$ & 1.46$\pm$0.02 & 1.77$\pm$0.20 & 53.2/73 &
 (2.5$\pm$0.1)$\times 10^{-8}$ & $18.42_{-0.44}^{+0.31}$ & $4.71_{-0.06}^{+0.10}$ & $<$0.16 & 323.2/73 \\
40031-03-02-00 & 2002-04-19 02:12-02:46 & (2.7$\pm$0.1)$\times 10^{-8}$
 & $63.0_{-2.6}^{+2.8}$ & 1.47$\pm$0.02 & 1.75$\pm$0.19 & 62.1/73 &
 $2.5_{-0.05}^{+0.04} \times 10^{-8}$	& $18.88_{-0.47}^{+0.10}$ & $4.66_{-0.03}^{+0.09}$ & $<$0.10 & 359.5/73 \\
40031-03-02-01 & 2002-04-20 02:55-04:09 & (2.8$\pm$0.1)$\times 10^{-8}$
 & $58.8_{-2.1}^{+2.3}$ & 1.45$\pm$0.02 & 1.97$\pm$0.19 & 69.4/73 &
 $2.6_{-0.01}^{+0.04} \times 10^{-8}$	& $19.46_{-0.42}^{+0.12}$ & $4.58_{-0.02}^{+0.04}$ & $<$0.07 & 357.1/73 \\
40031-03-02-02 & 2002-04-20 22:24-00:37 & (2.7$\pm$0.1)$\times 10^{-8}$
 & $58.1_{-1.7}^{+1.8}$ & 1.44$\pm$0.02 & $2.15_{-0.15}^{+0.16}$ &
 93.1/73 & $2.5_{-0.04}^{+0.05} \times 10^{-8}$ & $19.20_{-0.29}^{+0.24}$ & $4.64_{-0.05}^{+0.06}$ & $<$0.13 & 542.0/73 \\
70110-01-05-00 & 2002-04-21 11:09-11:24 & (2.7$\pm$0.1)$\times 10^{-8}$
 & $60.8_{-3.1}^{+3.4}$ & 1.46$\pm$0.02 & 1.86$\pm$0.22 & 65.1/73 &
 (2.6$\pm$0.1)$\times 10^{-8}$ & $18.33_{-0.50}^{+0.53}$ & 4.91$\pm$0.12 & 0.28$\pm$0.19 & 243.2/73 \\
40031-03-02-03 & 2002-04-21 22:07-03:07 & (2.7$\pm$0.1)$\times 10^{-8}$
 & $60.9_{-1.4}^{+1.5}$ & 1.47$\pm$0.01 & 1.96$\pm$0.13 & 92.9/73 &
 (2.6$\pm$0.4)$\times 10^{-8}$ & $20.28_{-0.22}^{+0.25}$ & $4.38_{-0.05}^{+0.04}$ & $<$0.03 & 898.7/73 \\
40031-03-02-04 & 2002-04-22 21:51-23:07 & (2.8$\pm$0.1)$\times 10^{-8}$
 & $60.1_{-2.1}^{+2.2}$ & 1.47$\pm$0.02 & 1.91$\pm$0.17 & 74.9/73 &
 $2.6_{-0.04}^{+0.05} \times 10^{-8}$	& $19.04_{-0.36}^{+0.18}$ & $4.65_{-0.04}^{+0.08}$ & $<$0.11 & 403.3/73 \\
70110-01-06-00 & 2002-04-23 11:39-11:54 & (2.8$\pm$0.1)$\times 10^{-8}$
 & $67.6_{-5.2}^{+6.0}$ & 1.50$\pm$0.03 & 1.67$\pm$0.27 & 62.4/73 &
 (2.7$\pm$0.1)$\times 10^{-8}$ & $18.50_{-0.74}^{+0.80}$ & 4.93$\pm$0.16 & 0.52$\pm$0.23 & 132.9/73 \\
70109-01-04-00 & 2002-04-23 13:20-18:13 & (2.7$\pm$0.1)$\times 10^{-8}$
 & $58.4_{-1.2}^{+1.2}$ & $1.47_{-0.01}^{+0.01}$ &
 $1.97_{-0.12}^{+0.12}$ & 127.7/73 & $2.6_{-0.04}^{+0.03} \times
 10^{-8}$ & $20.09_{-0.21}^{+0.21}$ & $4.37_{-0.04}^{+0.04}$ & $<$0.03 & 1190.1/73 \\
40031-03-02-05 & 2002-04-24 01:19-01:48 & (2.8$\pm$0.1)$\times 10^{-8}$
 & $60.1_{-2.7}^{+2.9}$ & 1.48$\pm$0.02 & 1.83$\pm$0.20 & 68.0/73 &
 (2.6$\pm$0.1)$\times 10^{-8}$ & $18.58_{-0.44}^{+0.47}$ & 4.77$\pm$0.10 & $<$0.17 & 315.8/73 \\
40031-03-02-06 & 2002-04-25 01:33-01:59 & (2.7$\pm$0.2)$\times 10^{-8}$
 & $59.5_{-5.9}^{+7.0}$ & 1.46$\pm$0.04 & 1.85$\pm$0.36 & 60.6/73 &
 (2.7$\pm$0.1)$\times 10^{-8}$ & $17.14_{-0.92}^{+1.02}$ & 5.26$\pm$0.24 & $0.74_{-0.31}^{+0.32}$ & 106.2/73 \\
70110-01-07-00 & 2002-04-26 21:43-22:20 & (2.7$\pm$0.1)$\times
 10^{-8}$  & $52.4_{-3.6}^{+4.1}$ & 1.46$\pm$0.03 & 1.90$\pm$0.27 &
 55.4/73 & (2.7$\pm$0.1)$\times 10^{-8}$ & $16.05_{-0.60}^{+0.65}$ & 5.31$\pm$0.17 & 0.65$\pm$0.24 & 142.7/73 \\
70109-01-05-01G & 2002-04-27 07:39-08:50 & (2.8$\pm$0.1)$\times 10^{-8}$ & $61.1_{-2.2}^{+2.4}$ & 1.50$\pm$0.02 & 1.72$\pm$0.18 & 47.3/73 & (2.6$\pm$0.1)$\times 10^{-8}$ & 19.14$\pm$0.34 & 4.60$\pm$0.07 & $<$0.06 & 398.4/73 \\
70109-01-05-00 & 2002-04-27 09:12-10:36 & (2.7$\pm$0.1)$\times 10^{-8}$
 & $60.8_{-2.2}^{+2.3}$ & 1.50$\pm$0.02 &
 $1.73_{-0.17}^{+0.18}$ & 71.1/73 & (2.6$\pm$0.1)$\times 10^{-8}$ & $19.12_{-0.32}^{+0.35}$ & $4.58_{-0.07}^{+0.07}$ & $<$0.06 & 434.9/73 \\
70109-01-05-02 & 2002-04-27 10:48-11:51 & (2.8$\pm$0.1)$\times 10^{-8}$
 & $60.3_{-2.1}^{+2.2}$ & 1.50$\pm$0.02 &
 $1.71_{-0.17}^{+0.17}$ & 77.1/73 & (2.6$\pm$0.1)$\times 10^{-8}$ & $19.16_{-0.31}^{+0.34}$ & $4.55_{-0.07}^{+0.07}$ & $<$0.04 & 487.8/73 \\
70110-01-08-00 & 2002-04-30 10:28-11:02 & (2.8$\pm$0.1)$\times 10^{-8}$
 & $61.9_{-2.6}^{+2.8}$ & 1.53$\pm$0.02 & 1.72$\pm$0.19 & 78.0/73 &
 (2.6$\pm$0.1)$\times 10^{-8}$ & $19.08_{-0.43}^{+0.40}$ & $4.56_{-0.08}^{+0.09}$ & $<$0.19 & 304.4/73 \\
70110-01-01-10 & 2003-04-11 00:19-00:36 & (2.0$\pm$0.2)$\times 10^{-9}$
 & -- & 1.66$\pm$0.08 & $1.31_{-0.62}^{+0.63}$ & 65.6/74 &
 $1.9_{-0.1}^{+0.2} \times 10^{-9}$ & $23.34_{-5.04}^{+9.92}$ & $4.48_{-0.88}^{+0.73}$ & 1.22$\pm$0.58 & 62.0/74 \\
80116-02-02-00 & 2003-04-12 16:45-19:56 & (2.0$\pm$0.1)$\times 10^{-9}$
 & $>$211.8 & $1.65_{-0.05}^{+0.04}$ & 1.37$\pm$0.38 &
 56.9/74 & (1.9$\pm$0.1)$\times 10^{-9}$ & $32.57_{-4.65}^{+8.05}$ & $3.64_{-0.48}^{+0.39}$ & $0.94_{-0.30}^{+0.31}$ & 61.6/74 \\
80116-02-02-01 & 2003-04-13 00:55-01:32 & ($2.1_{-0.2}^{+0.1}$)$\times 10^{-9}$ & -- & $1.63_{-0.06}^{+0.03}$ & $1.51_{-0.32}^{+0.49}$ & 65.2/74 & (2.1$\pm$0.1)$\times 10^{-9}$ & $31.00_{-6.37}^{+17.83}$ & $3.97_{-0.95}^{+0.60}$ & $1.37_{-0.39}^{+0.40}$ & 62.7/74 \\
70128-02-03-00 & 2003-04-13 05:39-11:00 & (1.9$\pm$0.1)$\times 10^{-9}$
 & $229.9_{-55.6}^{+100.4}$ & 1.59$\pm$0.03 & 1.49$\pm$0.29 & 68.6/74 &
 (1.8$\pm$0.1)$\times 10^{-9}$	& $27.14_{-2.19}^{+2.71}$ & $4.18_{-0.25}^{+0.24}$ & 0.97$\pm$0.24 & 62.5/74 \\
70128-02-03-01 & 2003-04-13 14:40-19:35 & (2.0$\pm$0.1)$\times 10^{-9}$
 & $>$212.1 & 1.59$\pm$0.04 & 1.54$\pm$0.36 & 57.3/74 & (1.9$\pm$0.1)$\times 10^{-9}$	& $30.81_{-3.33}^{+4.80}$ & $3.92_{-0.35}^{+0.31}$ & 0.98$\pm$0.28 & 61.8/74 \\
70110-01-02-10 & 2003-04-14 03:48-04:02 & (1.9$\pm$0.2)$\times 10^{-9}$ & -- & $1.64_{-0.08}^{+0.02}$ & $1.40_{-0.54}^{+0.60}$ & 59.9/74 & (2.1$\pm$0.2)$\times 10^{-9}$ & -- & $1.90_{-0.10}^{+2.18}$ & $1.15_{-0.52}^{+0.53}$ & 61.4/74 \\
80116-02-02-02G & 2003-04-14 06:22-09:28 & (1.9$\pm$0.1)$\times 10^{-9}$ & $265.4_{-83.3}^{+205.2}$ & 1.59$\pm$0.04 & 1.39$\pm$0.35 & 64.4/74 & (1.8$\pm$0.1)$\times 10^{-9}$ & $26.78_{-2.63}^{+3.46}$ & $4.27_{-0.33}^{+0.31}$ & 0.99$\pm$0.29 & 55.2/74 \\
60705-01-61-00 & 2003-04-17 23:05-00:01 & (1.3$\pm$0.1)$\times 10^{-9}$
 & $186.2_{-72.3}^{+246.5}$ & 1.55$\pm$0.06 & 1.23$\pm$0.48 & 60.0/74 &
 (1.3$\pm$0.1)$\times 10^{-9}$	& $23.52_{-4.00}^{+6.29}$ & $4.73_{-0.65}^{+0.59}$ & $0.85_{-0.42}^{+0.43}$ & 64.0/74 \\
70110-01-03-10 & 2003-04-18 09:39-10:06 & (1.4$\pm$0.2)$\times 10^{-9}$
 & $>$74.3 & $1.48_{-0.11}^{+0.10}$ & 2.29$\pm$0.79 & 48.7/74 &
 (1.3$\pm$0.1)$\times 10^{-9}$ & $20.23_{-5.57}^{+11.91}$ & $5.44_{-1.32}^{+1.16}$ & $2.05_{-0.75}^{+0.77}$ & 50.6/74 \\
80132-01-05-00 & 2004-02-21 09:16-10:10 & (1.1$\pm$0.1)$\times 10^{-9}$ & -- & $1.46_{-0.07}^{+0.05}$ & $1.18_{-0.50}^{+0.53}$ & 73.7/74 & (1.8$\pm$0.1)$\times 10^{-9}$ & $32.48_{-5.99}^{+13.21}$ & $4.35_{-0.80}^{+0.59}$ & 0.64$\pm$0.44 & 78.9/74 \\
80132-01-05-01 & 2004-02-21 13:10-14:17 & $1.8_{-0.1}^{+0.2}$ $\times 10^{-9}$ & $>$170.8 & $1.43_{-0.06}^{+0.05}$ & 1.27$\pm$0.48 & 76.2/74 & (1.7$\pm$0.1)$\times 10^{-9}$ & $26.81_{-3.49}^{+4.96}$ & $4.97_{-0.51}^{+0.47}$ & $0.82_{-0.40}^{+0.41}$ & 73.6/74 \\
80132-01-06-00 & 2004-02-22 05:34-06:29 & (2.0$\pm$0.2)$\times 10^{-9}$ & $>$200.4 & $1.46_{-0.06}^{+0.04}$ & $0.89_{-0.43}^{+0.45}$ &
 68.4/74 & (1.9$\pm$0.1)$\times 10^{-9}$ & $27.39_{-3.60}^{+5.38}$ & $4.80_{-0.52}^{+0.46}$ & 0.46$\pm$0.38 & 68.7/74 \\
80132-01-06-01 & 2004-02-22 20:51-22:14 & (1.9$\pm$0.1)$\times 10^{-9}$ & $>$197.5 & $1.47_{-0.05}^{+0.04}$ & $1.08_{-0.41}^{+0.43}$ &
 56.0/74 & (1.8$\pm$0.1)$\times 10^{-9}$ & $30.60_{-5.24}^{+10.67}$ & $4.43_{-0.72}^{+0.54}$ & 0.55$\pm$0.36 & 66.5/74 \\
80132-01-07-01 & 2004-02-23 20:30-21:52 & $2.2_{-0.2}^{+0.1}$ $\times 10^{-9}$ & $>$242.7 & $1.49_{-0.05}^{+0.03}$ & $0.86_{-0.34}^{+0.40}$ & 54.6/74 & (2.3$\pm$0.1)$\times 10^{-9}$ & $>$72.08 & $2.22_{-0.07}^{+0.47}$ & $0.39_{-0.16}^{+0.33}$ & 63.3/74 \\
90109-01-01-00 & 2004-02-23 23:43-04:11 & (2.1$\pm$0.1)$\times 10^{-9}$ & $290.3_{-83.4}^{+183.2}$ &
 1.43$\pm$0.04 & 1.35$\pm$0.36 & 63.3/74 & (1.9$\pm$0.1)$\times 10^{-9}$ & $30.05_{-2.52}^{+3.26}$ & $4.46_{-0.30}^{+0.28}$ & 0.42$\pm$0.27 & 77.8/74 \\
80102-04-58-01 & 2004-02-24 14:18-14:46 & (2.2$\pm$0.2)$\times 10^{-9}$ & -- & $1.48_{-0.07}^{+0.05}$ & $0.71_{-0.53}^{+0.59}$ & 63.7/74 & (4.6$\pm$0.3)$\times 10^{-9}$ & $26.20_{-4.61}^{+7.95}$ & $4.93_{-0.75}^{+0.64}$ & $<$0.97 & 61.3/74 \\
80132-01-07-00 & 2004-02-26 17:56-20:32 & (2.5$\pm$0.2)$\times 10^{-9}$ & $197.4_{-49.0}^{+91.6}$ & 1.39$\pm$0.05 & 1.43$\pm$0.39 &
 65.5/74 & (2.3$\pm$0.1)$\times 10^{-9}$ & $25.28_{-2.00}^{+2.43}$ & $5.03_{-0.31}^{+0.30}$ & $0.58_{-0.30}^{+0.31}$ & 59.7/74 \\
80102-04-59-00 & 2004-02-27 10:37-11:15 & (2.9$\pm$0.3)$\times 10^{-9}$ & -- & $1.44_{-0.06}^{+0.03}$ &
 $1.43_{-0.60}^{+0.61}$ & 47.7/74 & $2.1_{-0.1}^{+0.2}$ $\times 10^{-9}$ & $>$25.65 & $4.38_{-1.96}^{+0.82}$ & $0.99_{-0.58}^{+0.59}$ & 47.6/74 \\
80132-01-08-00 & 2004-02-27 19:08-20:42 & (2.6$\pm$0.2)$\times 10^{-9}$ & $231.9_{-76.0}^{+196.2}$ &
 1.42$\pm$0.05 & 1.06$\pm$0.39 & 62.6/74 & (2.5$\pm$0.1)$\times 10^{-9}$ & $24.66_{-2.52}^{+3.26}$ & $5.11_{-0.39}^{+0.37}$ & 0.51$\pm$0.33 & 61.8/74 \\
\hline
\label{table_2}
\end{tabular}
\end{center}
\end{minipage}}
\end{table*}


\setlength{\tabcolsep}{2.4pt}
\begin{table*}
\rotatebox{90}{
\begin{minipage}[t]{\textwidth}
\scriptsize
\begin{center}
\caption{Results of spectral fitting (III).}
\begin{tabular}{c c c c c c c c c c c c}
\hline
\hline
 & & cut-off power-law & (CPL) & & & & thermal & Comptonization & (COMPST) & & \\
\hline
Observation ID & Observation Time & Flux (erg/sec/cm$^2$) & $E_{\rm{cut}}$ (keV) &
 photon index & MAX$\tau$ & ${\chi}^2$/d.o.f & Flux (erg/sec/cm$^2$) &
 k$T_{\rm{e}}$ (keV) & $\tau$ & MAX$\tau$ & ${\chi}^2$/d.o.f  \\
\hline
80132-01-08-01 & 2004-02-28 07:51-08:53 & (2.8$\pm$0.2)$\times 10^{-9}$ & $>$213.1 & 1.44$\pm$0.04 &
 1.10$\pm$0.36 & 56.3/74 & (2.7$\pm$0.1)$\times 10^{-9}$ & $29.06_{-2.83}^{+3.78}$ & $4.63_{-0.35}^{+0.32}$ & 0.46$\pm$0.29 & 64.8/74 \\
80132-01-08-02 & 2004-02-29 16:53-17:58 & (3.1$\pm$0.2)$\times 10^{-9}$ & $>$214.0 & $1.44_{-0.05}^{+0.04}$
 & $1.28_{-0.38}^{+0.39}$ & 66.8/74 & (3.0$\pm$0.1)$\times 10^{-9}$ & $28.74_{-2.96}^{+3.99}$ & $4.69_{-0.38}^{+0.35}$ & 0.69$\pm$0.31 & 65.4/74 \\
80102-04-60-00 & 2004-03-01 07:52-08:13 & (3.6$\pm$0.4) $\times 10^{-9}$ & -- & $1.41_{-0.07}^{+0.06}$ &
 $1.40_{-0.29}^{+0.59}$ & 57.6/74 & (3.0$\pm$0.2)$\times 10^{-9}$ & $28.65_{-5.36}^{+10.37}$ & $4.87_{-0.83}^{+0.68}$ & 0.89$\pm$0.51 & 62.6/74 \\
80132-01-09-00 & 2004-03-01 18:12-19:11 & (3.6$\pm$0.2)$\times 10^{-9}$ & $252.8_{-78.0}^{+184.9}$ &
 1.44$\pm$0.04 & 1.19$\pm$0.36 & 78.5/74 & (3.3$\pm$0.1)$\times 10^{-9}$ & $26.53_{-2.58}^{+3.37}$ & $4.80_{-0.36}^{+0.33}$ & $0.54_{-0.29}^{+0.30}$ & 89.2/74 \\
80132-01-09-01 & 2004-03-02 19:19-19:42 & (4.0$\pm$0.3)$\times 10^{-9}$ & $>$170.9 & 1.44$\pm$0.05 &
 1.19$\pm$0.41 & 71.0/74 & (3.8$\pm$0.2)$\times 10^{-9}$ & $28.66_{-3.35}^{+4.71}$ & $4.64_{-0.43}^{+0.39}$ & $0.50_{-0.33}^{+0.34}$ & 82.0/74 \\
80132-01-09-02 & 2004-03-03 07:55-08:31 & (3.6$\pm$0.3)$\times 10^{-9}$ & -- & $1.45_{-0.06}^{+0.04}$ &
 $1.11_{-0.43}^{+0.48}$ & 79.1/74 & (3.4$\pm$0.2)$\times 10^{-9}$ & $28.07_{-3.50}^{+5.04}$ & $4.83_{-0.49}^{+0.44}$ & $0.75_{-0.40}^{+0.41}$ & 70.6/74 \\
80132-01-09-03 & 2004-03-03 23:41-00:42 & (4.0$\pm$0.2)$\times 10^{-9}$ & $215.3_{-45.5}^{+74.8}$ &
 1.41$\pm$0.03 & 1.24$\pm$0.30 & 74.2/74 & (3.8$\pm$0.1)$\times 10^{-9}$ & $26.33_{-1.63}^{+1.91}$ & 4.86$\pm$0.23 & $0.44_{-0.24}^{+0.25}$ & 69.2/74 \\
80102-04-61-00 & 2004-03-04 07:34-07:57 & $4.1_{-0.3}^{+0.4}$ $\times 10^{-9}$ & $>$156.2 & 1.40$\pm$0.06 &
 1.21$\pm$0.47 & 84.8/74 & (3.5$\pm$0.2)$\times 10^{-9}$ & $26.36_{-3.36}^{+4.73}$ & $5.04_{-0.51}^{+0.47}$ & 0.63$\pm$0.40 & 90.6/74 \\
80132-01-10-00 & 2004-03-04 20:08-00:01 & (4.0$\pm$0.2)$\times 10^{-9}$ & $233.6_{-40.7}^{+60.8}$ &
 1.41$\pm$0.03 & 1.31$\pm$0.27 & 74.6/74 & (3.6$\pm$0.1)$\times 10^{-9}$
 & $29.69_{-1.50}^{+1.74}$ & $4.46_{-0.18}^{+0.17}$ & $<$0.36 & 97.9/74
 \\
80132-01-11-00 & 2004-03-05 00:01-00:19 & (4.0$\pm$0.3)$\times 10^{-9}$ & $168.1_{-45.9}^{+91.2}$ &
 1.40$\pm$0.05 & $1.24_{-0.45}^{+0.46}$ & 71.9/74 & (3.8$\pm$0.2)$\times
 10^{-9}$ & $23.67_{-2.22}^{+2.77}$ & $5.22_{-0.39}^{+0.37}$ &
 0.58$\pm$0.38 & 65.7/74 \\
80132-01-11-01 & 2004-03-06 00:32-01:32 & (4.7$\pm$0.2)$\times 10^{-9}$ & $323.5_{-85.7}^{+173.6}$ &
 1.45$\pm$0.03 & 1.13$\pm$0.29 & 50.1/74 & (4.4$\pm$0.1)$\times 10^{-9}$ & $30.93_{-2.24}^{+2.82}$ & $4.36_{-0.24}^{+0.23}$ & 0.33$\pm$0.23 & 70.3/74 \\
80132-01-11-02 & 2004-03-06 19:36-20:52 & (4.8$\pm$0.3)$\times 10^{-9}$ & $185.5_{-36.1}^{+56.0}$ &
 1.40$\pm$0.04 & 1.29$\pm$0.31 & 60.4/74 & (4.5$\pm$0.1)$\times 10^{-9}$ & $25.23_{-1.58}^{+1.83}$ & $4.97_{-0.24}^{+0.23}$ & 0.42$\pm$0.25 & 64.7/74 \\
80132-01-12-00 & 2004-03-07 22:14-22:57 & (5.1$\pm$0.3)$\times 10^{-9}$ & $249.6_{-56.1}^{+97.1}$ &
 1.44$\pm$0.03 & 1.23$\pm$0.30 & 68.6/74 & (4.8$\pm$0.1)$\times 10^{-9}$ & $28.15_{-1.83}^{+2.21}$ & $4.61_{-0.23}^{+0.22}$ & 0.41$\pm$0.24 & 76.9/74 \\
80102-04-62-00 & 2004-03-08 07:40-07:59 & (4.4$\pm$0.4)$\times 10^{-9}$ & $181.3_{-56.2}^{+128.7}$ & 1.45$\pm$0.06 & 0.94$\pm$0.47 &
 66.0/74 & (3.9$\pm$0.2)$\times 10^{-9}$ & $24.05_{-2.63}^{+3.43}$ & $5.03_{-0.43}^{+0.41}$ & $0.38_{-0.38}^{+0.40}$ & 64.0/74 \\
80132-01-12-01 & 2004-03-08 18:47-19:39 & (5.1$\pm$0.3)$\times 10^{-9}$ & $244.8_{-54.3}^{+93.7}$ &
 1.44$\pm$0.03 & 1.22$\pm$0.30 & 60.4/74 & (4.8$\pm$0.1)$\times 10^{-9}$ & $27.64_{-1.73}^{+2.07}$ & $4.67_{-0.23}^{+0.22}$ & 0.43$\pm$0.24 & 60.0/74 \\
80132-01-13-00 & 2004-03-09 23:03-00:01 & (5.6$\pm$0.3)$\times 10^{-9}$ & $203.5_{-34.9}^{+51.5}$ &
 1.41$\pm$0.03 & 1.54$\pm$0.27 & 53.3/74 & (5.2$\pm$0.1)$\times 10^{-9}$ & $26.94_{-1.41}^{+1.62}$ & 4.77$\pm$0.19 & 0.58$\pm$0.22 & 65.6/74 \\
80132-01-13-01 & 2004-03-11 00:16-01:14 & (6.0$\pm$0.3)$\times 10^{-9}$ & $212.8_{-37.0}^{+55.2}$ &
 1.41$\pm$0.03 & 1.51$\pm$0.27 & 76.4/74 & (5.6$\pm$0.1)$\times 10^{-9}$ & $27.71_{-1.48}^{+1.70}$ & 4.66$\pm$0.19 & 0.55$\pm$0.21 & 99.4/74 \\
80102-04-63-00 & 2004-03-11 03:57-04:08 & (5.5$\pm$0.4)$\times 10^{-9}$ & $247.2_{-84.3}^{+230.1}$ &
 1.45$\pm$0.05 & $1.21_{-0.43}^{+0.44}$ & 80.6/74 & (5.3$\pm$0.2)$\times 10^{-9}$ & $26.32_{-2.69}^{+3.52}$ & $4.83_{-0.39}^{+0.37}$ & $0.70_{-0.36}^{+0.37}$ & 72.7/74 \\
90118-01-01-00 & 2004-03-11 22:18-00:01 & (6.4$\pm$0.3)$\times 10^{-9}$
 & $226.6_{-37.8}^{+56.0}$ & 1.42$\pm$0.03 & 1.51$\pm$0.25 & 88.8/74 &
 (5.9$\pm$0.1)$\times 10^{-9}$	& $28.73_{-1.45}^{+1.66}$ & $4.55_{-0.18}^{+0.17}$ & 0.51$\pm$0.20 & 116.8/74 \\
90118-01-02-00 & 2004-03-13 16:58-17:39 & (6.9$\pm$0.3)$\times 10^{-9}$
 & $173.3_{-26.9}^{+37.7}$ & 1.40$\pm$0.03 & 1.66$\pm$0.28 & 74.9/74 &
 (6.4$\pm$0.2)$\times 10^{-9}$	& $25.73_{-1.31}^{+1.50}$ & 4.85$\pm$0.19 & 0.67$\pm$0.22 & 87.9/74 \\
80102-04-64-00 & 2004-03-14 15:16-15:39 & (6.9$\pm$0.4)$\times 10^{-9}$
 & $234.7_{-59.7}^{+112.6}$ & 1.46$\pm$0.04 & 1.25$\pm$0.35 & 84.9/74 &
 (6.5$\pm$0.2)$\times 10^{-9}$	& $26.90_{-2.07}^{+2.55}$ & $4.65_{-0.28}^{+0.27}$ & 0.57$\pm$0.28 & 81.9/74 \\
80132-01-14-00 & 2004-03-14 16:28-17:14 & (6.8$\pm$0.3)$\times 10^{-9}$
 & $175.8_{-25.5}^{+35.3}$ & 1.42$\pm$0.03 & $1.52_{-0.27}^{+0.26}$ &
 91.2/74 & (6.4$\pm$0.2)$\times 10^{-9}$ & $25.83_{-1.20}^{+1.34}$ & 4.77$\pm$0.17 & 0.54$\pm$0.21 & 81.3/74 \\
90118-01-03-00 & 2004-03-15 02:25-03:00 & (6.6$\pm$0.4)$\times 10^{-9}$
 & $200.1_{-39.3}^{+62.7}$ & $1.42_{-0.03}^{+0.04}$ & 1.54$\pm$0.31 &
 90.8/74& (6.2$\pm$0.2)$\times 10^{-9}$	& $25.76_{-1.56}^{+1.81}$ & 4.87$\pm$0.23 & 0.74$\pm$0.25 & 85.8/74 \\ 
90118-01-04-00 & 2004-03-15 19:49-21:12 & (7.0$\pm$0.3)$\times 10^{-9}$
 & $195.5_{-30.6}^{+43.4}$ & 1.43$\pm$0.03 & 1.35$\pm$0.26 & 68.7/74 &
 (6.5$\pm$0.2)$\times 10^{-9}$	& $27.28_{-1.36}^{+1.53}$ & $4.60_{-0.18}^{+0.17}$ & 0.38$\pm$0.21 & 81.6/74 \\
90118-01-05-00 & 2004-03-16 17:19-17:58 & (7.2$\pm$0.3)$\times 10^{-9}$
 & $165.4_{-23.8}^{+32.5}$ & 1.42$\pm$0.03 & 1.44$\pm$0.27 & 72.0/74 &
 (6.7$\pm$0.2)$\times 10^{-9}$ & $25.52_{-1.24}^{+1.39}$ & 4.80$\pm$0.18 & $0.45_{-0.21}^{+0.22}$ & 87.7/74 \\
90118-01-06-00 & 2004-03-17 12:13-13:00 & (7.3$\pm$0.4)$\times 10^{-9}$
 & $144.2_{-21.5}^{+29.4}$ & 1.41$\pm$0.03 & 1.63$\pm$0.29 & 55.2/74 &
 (6.8$\pm$0.2)$\times 10^{-9}$	& $23.93_{-1.30}^{+1.47}$ & 4.97$\pm$0.21 & 0.67$\pm$0.24 & 66.8/74 \\
80102-04-65-00 & 2004-03-18 07:06-07:35 & (7.8$\pm$0.6)$\times 10^{-9}$
 & $209.2_{-59.0}^{+124.8}$ & 1.48$\pm$0.05 & 1.24$\pm$0.42 & 87.8/74 &
 (7.5$\pm$0.3)$\times 10^{-9}$	& $25.27_{-2.25}^{+2.83}$ & $4.80_{-0.34}^{+0.33}$ & 0.68$\pm$0.35 & 76.9/74 \\
90118-01-07-00 & 2004-03-18 18:08-20:59 & (7.8$\pm$0.3)$\times 10^{-9}$
 & $163.0_{-15.5}^{+18.7}$ & 1.42$\pm$0.02 & 1.64$\pm$0.21 & 79.3/74 &
 $7.2_{-0.1}^{+0.2} \times 10^{-9}$ & $26.85_{-0.83}^{+0.90}$ & 4.57$\pm$0.11 & $0.37_{-0.16}^{+0.17}$ & 90.6/74 \\
80132-01-15-00 & 2004-03-19 15:34-17:09 & (8.0$\pm$0.3)$\times 10^{-9}$
 & $168.4_{-20.2}^{+26.2}$ & 1.42$\pm$0.03 & 1.66$\pm$0.24 & 103.1/74 &
 (7.4$\pm$0.2)$\times 10^{-9}$	& $25.99_{-1.00}^{+1.10}$ & 4.71$\pm$0.14 & 0.59$\pm$0.19 & 80.8/74 \\
80102-04-66-00 & 2004-03-20 11:05-11:47 & (7.8$\pm$0.4)$\times 10^{-9}$
 & $174.5_{-29.8}^{+43.7}$ & 1.43$\pm$0.03 & 1.65$\pm$0.29 & 91.8/74 &
 (7.3$\pm$0.2)$\times 10^{-9}$	& $25.18_{-1.46}^{+1.70}$ & $4.83_{-0.22}^{+0.21}$ & 0.76$\pm$0.24 & 104.0/74 \\
80132-01-15-02 & 2004-03-22 13:29-14:28 & (8.3$\pm$0.3)$\times 10^{-9}$
 & $156.5_{-17.6}^{+22.4}$ & 1.42$\pm$0.03 & 1.73$\pm$0.24 & 85.2/74 &
 (7.7$\pm$0.2)$\times 10^{-9}$	& $25.61_{-0.97}^{+1.07}$ & 4.72$\pm$0.14 & 0.61$\pm$0.19 & 87.9/74 \\
80132-01-15-03 & 2004-03-23 21:15-21:58 & (8.4$\pm$0.4)$\times 10^{-9}$
 & $150.2_{-17.8}^{+22.5}$ & 1.41$\pm$0.03 & 1.81$\pm$0.25 & 79.1/74 &
 (7.8$\pm$0.2)$\times 10^{-9}$	& $25.37_{-1.04}^{+1.15}$ & $4.76_{-0.16}^{+0.15}$ & 0.70$\pm$0.20 & 94.1/74 \\
80102-04-67-00 & 2004-03-24 03:53-04:09 & (8.4$\pm$0.5)$\times 10^{-9}$
 & $183.1_{-39.1}^{+64.1}$ & 1.46$\pm$0.04 & $1.30_{-0.32}^{+0.33}$ &
 84.3/74 & (7.9$\pm$0.3)$\times 10^{-9}$ & $25.24_{-1.88}^{+2.26}$ & $4.76_{-0.27}^{+0.26}$ & 0.55$\pm$0.27 & 93.7/74 \\
80132-01-15-01 & 2004-03-25 16:02-16:37 & (8.1$\pm$0.4)$\times 10^{-9}$
 & $161.2_{-24.1}^{+33.7}$ & 1.44$\pm$0.03 & 1.44$\pm$0.28 & 81.9/74 &
 (7.6$\pm$0.2)$\times 10^{-9}$	& $25.22_{-1.31}^{+1.50}$ & 4.75$\pm$0.19 & 0.52$\pm$0.23 & 88.2/74 \\
80132-01-16-00 & 2004-03-26 11:58-13:50 & (8.5$\pm$0.3)$\times 10^{-9}$
 & $155.6_{-16.2}^{+20.1}$ & 1.41$\pm$0.02 & 1.77$\pm$0.22 & 93.9/74 &
 (7.9$\pm$0.2)$\times 10^{-9}$	& $25.71_{-0.89}^{+0.97}$ & 4.74$\pm$0.13 & 0.59$\pm$0.18 & 95.4/74 \\
80132-01-16-01 & 2004-03-27 18:19-18:54 & (8.4$\pm$0.4)$\times 10^{-9}$
 & $153.7_{-20.1}^{+26.3}$ & 1.43$\pm$0.03 &
 $1.71_{-0.27}^{+0.27}$ & 53.4/74 & (7.9$\pm$0.2)$\times 10^{-9}$ & $25.47_{-1.18}^{+1.32}$ & 4.72$\pm$0.17 & 0.66$\pm$0.21 & 76.9/74 \\
80132-01-17-00 & 2004-03-28 11:14-12:45 & (8.3$\pm$0.3)$\times 10^{-9}$
 & $152.3_{-16.7}^{+21.2}$ & 1.42$\pm$0.02 & 1.67$\pm$0.23 & 70.0/74 &
 (7.8$\pm$0.2)$\times 10^{-9}$	& $25.79_{-1.00}^{+1.10}$ & 4.66$\pm$0.14 & 0.51$\pm$0.19 & 98.8/74 \\
80102-04-69-00 & 2004-03-29 08:23-08:50 &  (8.6$\pm$0.5)$\times
 10^{-9}$ & $167.9_{-30.8}^{+46.7}$ & 1.44$\pm$0.04 & 1.69$\pm$0.33 &
 66.7/74 & (8.1$\pm$0.3)$\times 10^{-9}$ & $24.94_{-1.57}^{+1.85}$ & 4.82$\pm$0.24 & 0.87$\pm$0.27 & 68.5/74 \\
80132-01-17-01 & 2004-03-29 10:52-12:22 & $8.5_{-0.3}^{+0.4} \times
 10^{-9}$ & $168.2_{-19.7}^{+25.4}$ & 1.44$\pm$0.02 & 1.66$\pm$0.23 &
 71.8/74 & (7.9$\pm$0.2)$\times 10^{-9}$ & $26.50_{-1.05}^{+1.16}$ & 4.57$\pm$0.14 & 0.58$\pm$0.19 & 89.9/74 \\
80132-01-18-00 & 2004-03-30 07:19-07:48 & $8.1_{-0.4}^{+0.5} \times
 10^{-9}$  & $129.5_{-19.5}^{+26.7}$ & 1.40$\pm$0.04 &
 $1.78_{-0.31}^{+0.32}$ & 71.2/74 & $7.6_{-0.2}^{+0.3} \times 10^{-9}$	& $22.69_{-1.29}^{+1.46}$ & 5.13$\pm$0.23 & 0.87$\pm$0.26 & 74.4/74 \\
80102-04-68-00 & 2004-03-31 07:37-08:17 & $8.7_{-0.4}^{+0.5} \times
 10^{-9}$ & $178.6_{-30.3}^{+44.6}$ & 1.45$\pm$0.03 & 1.55$\pm$0.30 &
 73.0/74 & (8.1$\pm$0.2)$\times 10^{-9}$ & $25.65_{-1.44}^{+1.66}$ & $4.72_{-0.21}^{+0.20}$ & 0.70$\pm$0.24 & 76.0/74 \\
\hline
\label{table_3}
\end{tabular}
\end{center}
\end{minipage}}
\end{table*}


\setlength{\tabcolsep}{2.4pt}
\begin{table*}
\rotatebox{90}{
\begin{minipage}[t]{\textwidth}
\scriptsize
\begin{center}
\caption{Results of spectral fitting (IV).}
\begin{tabular}{c c c c c c c c c c c c}
\hline
\hline
 & & cut-off power-law & (CPL) & & & & thermal & Comptonization & (COMPST) & & \\
\hline
Observation ID & Observation Time & Flux (erg/sec/cm$^2$) & $E_{\rm{cut}}$ (keV) &
 photon index & MAX$\tau$ & ${\chi}^2$/d.o.f & Flux (erg/sec/cm$^2$) &
 k$T_{\rm{e}}$ (keV) & $\tau$ & MAX$\tau$ & ${\chi}^2$/d.o.f  \\
\hline
80132-01-18-01 & 2004-03-31 15:30-16:43 & $8.5_{-0.4}^{+0.5} \times
 10^{-9}$ & $185.5_{-31.5}^{+46.2}$ & 1.47$\pm$0.03 & 1.62$\pm$0.30 &
 73.9/74 & (7.9$\pm$0.2)$\times 10^{-9}$ & $26.89_{-1.54}^{+1.82}$ & $4.50_{-0.21}^{+0.20}$ & 0.66$\pm$0.24 & 84.3/74 \\
80132-01-18-02 & 2004-04-01 08:06-09:12 & $8.6_{-0.3}^{+0.4} \times
 10^{-9}$ 	& $147.0_{-15.4}^{+18.8}$ & 1.43$\pm$0.02 &
 1.68$\pm$0.23 & 79.5/74 & (8.0$\pm$0.2)$\times 10^{-9}$ & $25.61_{-0.97}^{+1.07}$ & 4.63$\pm$0.14 & 0.49$\pm$0.18 & 141.2/74 \\
80132-01-19-00 & 2004-04-02 06:54-07:25 & (6.4$\pm$0.3)$\times 10^{-9}$
 & $148.7_{-22.4}^{+30.5}$ & 1.42$\pm$0.03 & 1.39$\pm$0.28 & 76.3/74 &
 (6.1$\pm$0.2)$\times 10^{-9}$	& $23.76_{-1.24}^{+1.39}$ & 4.99$\pm$0.20 & 0.55$\pm$0.23 & 76.7/74 \\
80132-01-19-01 & 2004-04-03 10:36-12:00 & $8.7_{-0.3}^{+0.4} \times
 10^{-9}$ 	& $161.4_{-18.0}^{+22.5}$ & 1.44$\pm$0.02 &
 1.74$\pm$0.23 & 65.9/74 & (8.1$\pm$0.2)$\times 10^{-9}$ & $26.28_{-0.98}^{+1.08}$ & $4.59_{-0.14}^{+0.13}$ & 0.63$\pm$0.18 & 64.0/74 \\
80132-01-19-05 & 2004-04-04 02:50-03:09 & (8.6$\pm$0.5)$\times 10^{-9}$
 & $195.3_{-38.9}^{+61.6}$ & 1.46$\pm$0.04 & 1.60$\pm$0.31 & 59.8/74 &
 $8.0_{-0.2}^{+0.3} \times 10^{-9}$ & $25.56_{-1.66}^{+1.97}$ & $4.73_{-0.24}^{+0.23}$ & 0.83$\pm$0.25 & 68.7/74 \\
80102-04-70-00 & 2004-04-04 14:12-14:47 & $9.0_{-0.6}^{+0.7} \times
 10^{-9}$ 	& $>$190.6 & 1.46$\pm$0.05 & 1.52$\pm$0.40 & 61.4/74 &
 $8.6_{-0.3}^{+0.4} \times 10^{-9}$ & $27.50_{-2.55}^{+3.31}$ & $4.71_{-0.35}^{+0.33}$ & 0.98$\pm$0.33 & 56.7/74 \\
80132-01-19-02 & 2004-04-05 21:07-22:43 & (8.4$\pm$0.4)$\times 10^{-9}$
 & $168.6_{-22.8}^{+30.4}$ & 1.44$\pm$0.03 & 1.68$\pm$0.25 & 78.4/74 &
 (7.8$\pm$0.2)$\times 10^{-9}$	& $25.75_{-1.16}^{+1.31}$ & 4.66$\pm$0.16 & 0.68$\pm$0.20 & 82.8/74 \\
80132-01-19-03 & 2004-04-06 18:10-19:00 & (8.5$\pm$0.5)$\times 10^{-9}$
 & $142.0_{-23.0}^{+32.4}$ & 1.42$\pm$0.04 & 1.73$\pm$0.33 & 85.0/74 &
 (8.1$\pm$0.3)$\times 10^{-9}$	& $23.44_{-1.35}^{+1.54}$ & 5.01$\pm$0.23 & 0.89$\pm$0.27 & 72.3/74 \\
80102-04-71-00 & 2004-04-07 04:45-05:01 & $8.2_{-0.4}^{+0.5} \times
 10^{-9}$ & $159.8_{-27.7}^{+41.2}$ & 1.44$\pm$0.04 & 1.66$\pm$0.32 &
 59.5/74 & $7.7_{-0.2}^{+0.3} \times 10^{-9}$ & $24.78_{-1.54}^{+1.79}$ & 4.81$\pm$0.23 & 0.80$\pm$0.26 & 70.3/74 \\
80132-01-20-02 & 2004-04-09 11:10-11:28 & (7.8$\pm$0.5)$\times 10^{-9}$
 & $147.8_{-27.4}^{+41.6}$ & 1.43$\pm$0.04 & 1.65$\pm$0.34 & 66.0/74 &
 (7.3$\pm$0.3)$\times 10^{-9}$	& $23.32_{-1.63}^{+1.93}$ & $5.00_{-0.28}^{+0.27}$ & 0.85$\pm$0.29 & 80.1/74 \\
80132-01-20-04 & 2004-04-10 07:40-07:58 & (7.9$\pm$0.5)$\times 10^{-9}$
 & $226.2_{-55.5}^{+103.8}$ & 1.48$\pm$0.04 & 1.32$\pm$0.33 & 74.3/74 &
 (7.5$\pm$0.3)$\times 10^{-9}$	& $26.47_{-2.01}^{+2.44}$ & $4.65_{-0.27}^{+0.26}$ & 0.69$\pm$0.27 & 71.3/74 \\
80132-01-20-00 & 2004-04-10 10:49-11:07 & (7.6$\pm$0.5)$\times 10^{-9}$ & $207.0_{-49.6}^{+88.0}$ & 1.48$\pm$0.04 & 1.39$\pm$0.34 & 69.6/74 & (7.2$\pm$0.3)$\times 10^{-9}$ & $25.19_{-1.86}^{+2.24}$ & 4.77$\pm$0.27 & 0.82$\pm$0.28 & 59.8/74 \\
80132-01-20-05 & 2004-04-10 09:15-09:32 & (7.6$\pm$0.5)$\times 10^{-9}$
 & $156.3_{-32.3}^{+51.8}$ & 1.45$\pm$0.04 & 1.43$\pm$0.34 & 52.4/74 &
 (7.2$\pm$0.3)$\times 10^{-9}$	& $24.60_{-2.05}^{+2.48}$ & 4.79$\pm$0.30 & 0.65$\pm$0.29 & 80.2/74 \\
80132-01-20-06 & 2004-04-11 04:10-04:28 & $7.6_{-0.4}^{+0.5} \times
 10^{-9}$ 	& $178.4_{-37.2}^{+60.0}$ & 1.46$\pm$0.04 &
 1.56$\pm$0.34 & 90.6/74 & $7.2_{-0.2}^{+0.3} \times 10^{-9}$	& $25.17_{-1.82}^{+2.17}$ & $4.76_{-0.27}^{+0.26}$ & 0.82$\pm$0.28 & 98.2/74 \\
80132-01-20-07 & 2004-04-11 05:45-06:02 & (7.7$\pm$0.5)$\times 10^{-9}$
 & $230.7_{-60.3}^{+120.2}$ & 1.48$\pm$0.04 & 1.25$\pm$0.35 & 68.4/74 &
 (7.3$\pm$0.3)$\times 10^{-9}$	& $26.91_{-2.36}^{+3.06}$ & $4.58_{-0.32}^{+0.30}$ & 0.59$\pm$0.29 & 85.5/74 \\
80132-01-20-08 & 2004-04-11 08:54-09:12 & (7.9$\pm$0.5)$\times 10^{-9}$
 & $173.6_{-36.4}^{+59.0}$ & 1.45$\pm$0.04 & 1.67$\pm$0.34 & 74.4/74 &
 (7.5$\pm$0.3)$\times 10^{-9}$	& $25.04_{-1.90}^{+2.29}$ & $4.79_{-0.28}^{+0.27}$ & 0.92$\pm$0.28 & 88.4/74 \\
80132-01-20-09 & 2004-04-11 10:29-10:46 & (7.7$\pm$0.5)$\times 10^{-9}$
 & $171.1_{-36.8}^{+60.1}$ & 1.45$\pm$0.04 & $1.55_{-0.34}^{+0.35}$ &
 66.6/74 & (7.3$\pm$0.3)$\times 10^{-9}$ & $24.99_{-1.92}^{+2.32}$ & $4.82_{-0.29}^{+0.28}$ & 0.80$\pm$0.29 & 77.8/74 \\
80132-01-20-10 & 2004-04-12 02:15-02:33 & $7.5_{-0.4}^{+0.5} \times
 10^{-9}$ 	& $186.5_{-39.8}^{+65.0}$ & 1.46$\pm$0.04 &
 1.41$\pm$0.33 & 99.5/74 & (7.0$\pm$0.2)$\times 10^{-9}$ & $24.53_{-1.71}^{+2.02}$ & 4.86$\pm$0.26 & $0.75_{-0.27}^{+0.28}$ & 91.1/74 \\
80132-01-20-01 & 2004-04-12 10:01-10:26 & (6.9$\pm$0.7)$\times 10^{-9}$
 & $178.6_{-60.1}^{+158.4}$ & 1.51$\pm$0.07 & $1.21_{-0.54}^{+0.55}$ &
 79.6/74 & (6.6$\pm$0.4)$\times 10^{-9}$ & $22.61_{-2.70}^{+3.59}$ & $5.01_{-0.48}^{+0.47}$ & $0.85_{-0.47}^{+0.48}$ & 68.8/74 \\
80132-01-20-03 & 2004-04-13 03:30-03:47 & (7.4$\pm$0.5)$\times 10^{-9}$
 & $165.4_{-32.0}^{+50.0}$ & 1.47$\pm$0.04 & 1.61$\pm$0.38 & 78.9/74 &
 (7.0$\pm$0.3)$\times 10^{-9}$	& $25.16_{-1.68}^{+1.99}$ & $4.66_{-0.26}^{+0.25}$ & 0.75$\pm$0.30 & 66.8/74 \\
80102-04-73-00 & 2004-04-13 09:34-10:06 & ($7.7_{-0.6}^{+0.5}$)$\times 10^{-9}$ & -- & $1.51_{-0.07}^{+0.04}$ & $0.94_{-0.47}^{+0.57}$ & 74.4/74 & (4.1$\pm$0.3)$\times 10^{-9}$ & $30.17_{-4.81}^{+8.58}$ & $4.41_{-0.66}^{+0.55}$ & 0.63$\pm$0.51 & 69.3/74 \\
80132-01-20-11 & 2004-04-15 04:39-04:56 & (7.1$\pm$0.5)$\times 10^{-9}$
 & $>$178.6 & 1.49$\pm$0.05 & 1.26$\pm$0.41 & 65.4/74 & (6.7$\pm$0.3)$\times 10^{-9}$	& $27.91_{-3.32}^{+4.86}$ & $4.50_{-0.45}^{+0.40}$ & 0.71$\pm$0.35 & 75.9/74 \\
80102-04-74-00 & 2004-04-17 15:06-15:21 & $6.9_{-0.4}^{+0.5} \times
 10^{-9}$ 	& $167.8_{-38.6}^{+67.1}$ & 1.43$\pm$0.04 &
 1.57$\pm$0.37 & 50.1/74 & $6.5_{-0.2}^{+0.3} \times 10^{-9}$ & $23.65_{-1.91}^{+2.32}$ & 5.06$\pm$0.31 & $0.87_{-0.30}^{+0.31}$ & 52.1/74 \\
80132-01-21-00 & 2004-04-18 18:11-19:23 & (6.5$\pm$0.3)$\times 10^{-9}$
 & $196.2_{-35.7}^{+53.3}$ & 1.45$\pm$0.03 & 1.51$\pm$0.28 & 64.7/74 &
 (6.1$\pm$0.2)$\times 10^{-9}$	& $26.18_{-1.49}^{+1.72}$ & 4.69$\pm$0.20 & 0.68$\pm$0.22 & 60.4/74 \\
80132-01-21-01 & 2004-04-19 11:12-12:19 & (6.3$\pm$0.3)$\times 10^{-9}$
 & $200.9_{-38.4}^{+59.7}$ & 1.46$\pm$0.03 & $1.40_{-0.29}^{+0.30}$ &
 61.1/74 & (5.9$\pm$0.2)$\times 10^{-9}$ & $26.59_{-1.60}^{+1.86}$ & 4.61$\pm$0.21 & 0.62$\pm$0.24 & 55.3/74 \\
80132-01-21-02 & 2004-04-20 12:40-13:10 & (6.3$\pm$0.4)$\times 10^{-9}$
 & $230.1_{-57.0}^{+106.8}$ & 1.48$\pm$0.04 & 1.38$\pm$0.34 & 71.1/74 &
 (5.9$\pm$0.2)$\times 10^{-9}$	& $26.44_{-1.97}^{+2.43}$ & $4.66_{-0.27}^{+0.26}$ & 0.74$\pm$0.28 & 63.7/74 \\
80102-04-75-00 & 2004-04-21 10:43-11:23 & (6.0$\pm$0.4)$\times 10^{-9}$
 & $>$195.1 & 1.50$\pm$0.04 & 1.09$\pm$0.37 & 77.7/74 &
 (5.6$\pm$0.2)$\times 10^{-9}$ & $27.41_{-2.42}^{+3.10}$ & $4.56_{-0.32}^{+0.30}$ & $0.57_{-0.30}^{+0.31}$ & 69.4/74 \\
80132-01-21-03 & 2004-04-21 17:10-18:17 & (6.0$\pm$0.3)$\times 10^{-9}$
 & $253.1_{-62.4}^{+118.0}$ & 1.47$\pm$0.04 & 1.42$\pm$0.31 & 86.3/74 &
 (5.6$\pm$0.2)$\times 10^{-9}$	& $27.71_{-1.98}^{+2.42}$ & $4.55_{-0.25}^{+0.24}$ & 0.74$\pm$0.25 & 84.0/74 \\
90019-01-01-00 & 2004-04-22 03:49-04:51 & (5.9$\pm$0.3)$\times 10^{-9}$
 & $208.7_{-36.9}^{+55.1}$ & 1.45$\pm$0.03 & 1.38$\pm$0.27 & 64.5/74 &
 (5.5$\pm$0.1)$\times 10^{-9}$	& $27.67_{-1.58}^{+1.84}$ & $4.53_{-0.20}^{+0.19}$ & 0.49$\pm$0.21 & 89.4/74 \\
80132-01-21-04 & 2004-04-22 11:57-12:36 & (5.8$\pm$0.4)$\times 10^{-9}$
 & $>$215.5 & 1.52$\pm$0.04 & 1.18$\pm$0.37 & 72.1/74 &
 (5.5$\pm$0.2)$\times 10^{-9}$ & $28.66_{-2.75}^{+3.63}$ & $4.40_{-0.33}^{+0.31}$ & 0.68$\pm$0.30 & 66.2/74 \\
90019-01-02-00 & 2004-04-23 03:26-04:34 & (5.5$\pm$0.3)$\times 10^{-9}$
 & $202.7_{-36.0}^{+54.1}$ & 1.46$\pm$0.03 & 1.59$\pm$0.27 & 76.5/74 &
 (5.2$\pm$0.1)$\times 10^{-9}$	& $26.46_{-1.43}^{+1.65}$ & 4.66$\pm$0.19 & 0.79$\pm$0.22 & 66.0/74 \\
80102-04-76-00 & 2004-04-23 19:13-19:26 & (5.2$\pm$0.4)$\times 10^{-9}$
 & $182.3_{-49.8}^{+99.7}$ & 1.47$\pm$0.05 & 1.37$\pm$0.41 & 47.2/74 &
 (4.9$\pm$0.2)$\times 10^{-9}$	& $25.00_{-2.58}^{+3.36}$ & $4.79_{-0.39}^{+0.37}$ & 0.71$\pm$0.35 & 57.9/74 \\
80132-01-22-00 & 2004-04-24 15:38-17:11 & (5.3$\pm$0.3)$\times 10^{-9}$
 & $222.1_{-46.7}^{+76.9}$ & 1.46$\pm$0.03 & 1.41$\pm$0.29 & 57.0/74 &
 (5.0$\pm$0.1)$\times 10^{-9}$	& $26.93_{-1.77}^{+2.10}$ & $4.62_{-0.23}^{+0.22}$ & 0.63$\pm$0.23 & 69.5/74 \\
80132-01-22-01 & 2004-04-26 19:53-20:39 & (5.0$\pm$0.3)$\times 10^{-9}$
 & $236.0_{-51.7}^{+88.2}$ & 1.46$\pm$0.03 & 1.46$\pm$0.30 & 56.2/74 &
 (4.7$\pm$0.1)$\times 10^{-9}$	& $27.21_{-1.74}^{+2.05}$ & $4.61_{-0.23}^{+0.22}$ & 0.73$\pm$0.24 & 49.3/74 \\
80102-04-77-00 & 2004-04-27 19:17-19:32 & $4.7_{-0.3}^{+0.4} \times
 10^{-9}$ & $179.4_{-46.0}^{+87.4}$ & 1.45$\pm$0.05 & 1.57$\pm$0.41 &
 65.8/74 & (4.4$\pm$0.2)$\times 10^{-9}$ & $23.63_{-1.99}^{+2.41}$ & $5.06_{-0.34}^{+0.33}$ & 1.01$\pm$0.35 & 51.8/74 \\
80132-01-22-02 & 2004-04-28 20:31-21:28 & (4.7$\pm$0.2)$\times 10^{-9}$
 & $>$239.6 & 1.49$\pm$0.03 & 1.24$\pm$0.29 & 78.6/74 & (4.4$\pm$0.1)$\times 10^{-9}$	& $29.39_{-2.04}^{+2.50}$ & $4.40_{-0.23}^{+0.22}$ & 0.60$\pm$0.23 & 74.9/74 \\
80132-01-23-00 & 2004-04-30 15:10-15:43 & (4.5$\pm$0.3)$\times 10^{-9}$
 & $>$210.2 & 1.48$\pm$0.04 & 1.29$\pm$0.34 & 72.6/74 & (4.2$\pm$0.1)$\times 10^{-9}$	& $28.88_{-2.70}^{+3.57}$ & $4.48_{-0.33}^{+0.30}$ & 0.66$\pm$0.28 & 84.2/74 \\
80102-04-78-00 & 2004-05-01 14:37-14:51 & $4.3_{-0.3}^{+0.4} \times
 10^{-9}$ & $>$183.7 & $1.48_{-0.06}^{+0.05}$ & 1.39$\pm$0.47 & 52.4/74
 & (4.0$\pm$0.2)$\times 10^{-9}$ & $26.51_{-3.18}^{+4.47}$ & $4.77_{-0.47}^{+0.43}$ & $0.93_{-0.39}^{+0.40}$ & 51.1/74 \\
80132-01-23-01 & 2004-05-02 19:02-19:57 & (4.1$\pm$0.2)$\times 10^{-9}$
 & $247.1_{-60.2}^{+111.7}$ & 1.47$\pm$0.04 & $1.24_{-0.31}^{+0.31}$ &
 58.8/74 & (3.9$\pm$0.1)$\times 10^{-9}$ & $28.24_{-2.13}^{+2.64}$ & $4.49_{-0.26}^{+0.25}$ & 0.51$\pm$0.25 & 70.5/74 \\
80102-04-79-00 & 2004-05-03 17:42-18:31 & (4.1$\pm$0.3)$\times 10^{-9}$
 & $204.8_{-60.0}^{+132.5}$ & 1.45$\pm$0.05 & 1.23$\pm$0.43 & 62.0/74 &
 (3.9$\pm$0.2)$\times 10^{-9}$	& $24.47_{-2.30}^{+2.89}$ & $5.00_{-0.37}^{+0.36}$ & 0.69$\pm$0.36 & 52.6/74 \\
\hline
\label{table_4}
\end{tabular}
\end{center}
\end{minipage}}
\end{table*}


\setlength{\tabcolsep}{2.4pt}
\begin{table*}
\rotatebox{90}{
\begin{minipage}[t]{\textwidth}
\scriptsize
\begin{center}
\caption{Results of spectral fitting (V).}
\begin{tabular}{c c c c c c c c c c c c}
\hline
\hline
 & & cut-off power-law & (CPL) & & & & thermal & Comptonization & (COMPST) & & \\
\hline
Observation ID & Observation Time & Flux (erg/sec/cm$^2$) & $E_{\rm{cut}}$ (keV) &
 photon index & MAX$\tau$ & ${\chi}^2$/d.o.f & Flux (erg/sec/cm$^2$) &
 k$T_{\rm{e}}$ (keV) & $\tau$ & MAX$\tau$ & ${\chi}^2$/d.o.f  \\
\hline
80132-01-23-02 & 2004-05-04 21:43-22:23 & (4.2$\pm$0.3)$\times 10^{-9}$
 & $>$212.7 & 1.48$\pm$0.04 & 1.17$\pm$0.35 & 48.7/74 & (4.0$\pm$0.1)$\times 10^{-9}$	& $29.45_{-2.99}^{+4.13}$ & $4.43_{-0.36}^{+0.32}$ & 0.54$\pm$0.28 & 64.1/74 \\
80132-01-23-03 & 2004-05-06 11:09-11:58 & $4.0_{-0.2}^{+0.3} \times
 10^{-9}$ 	& $178.7_{-40.7}^{+70.1}$ & 1.44$\pm$0.04 &
 1.36$\pm$0.34 & 65.5/74 & (3.7$\pm$0.1)$\times 10^{-9}$ & $25.15_{-2.06}^{+2.53}$ & $4.83_{-0.30}^{+0.29}$ & 0.61$\pm$0.28 & 77.0/74 \\
80102-04-80-00 & 2004-05-09 02:52-03:07 & (3.7$\pm$0.3)$\times 10^{-9}$
 & $>$140.7 & 1.52$\pm$0.06 & 1.02$\pm$0.48 & 48.4/74 & (3.5$\pm$0.2)$\times 10^{-9}$	& $25.04_{-3.97}^{+6.35}$ & $4.74_{-0.62}^{+0.54}$ & 0.62$\pm$0.42 & 50.7/74 \\
90019-01-03-00 & 2004-05-11 09:17-09:52 & (4.1$\pm$0.3)$\times 10^{-9}$
 & $>$205.5 & 1.48$\pm$0.04 & 1.30$\pm$0.36 & 57.3/74 & (3.9$\pm$0.1)$\times 10^{-9}$	& $28.90_{-2.99}^{+4.09}$ & $4.50_{-0.37}^{+0.34}$ & 0.70$\pm$0.29 & 67.6/74 \\
90019-01-03-01 & 2004-05-12 08:53-09:30 & $3.8_{-0.2}^{+0.3} \times
 10^{-9}$ & $261.0_{-79.8}^{+192.9}$ & 1.47$\pm$0.04 & 1.28$\pm$0.36 &
 61.9/74 & (3.6$\pm$0.1)$\times 10^{-9}$ & $25.72_{-2.26}^{+2.82}$ & $4.83_{-0.32}^{+0.31}$ & 0.78$\pm$0.30 & 50.8/74 \\
90019-01-04-00 & 2004-05-14 08:08-08:39 & (3.9$\pm$0.3)$\times 10^{-9}$
 & $253.6_{-79.2}^{+190.4}$ & 1.45$\pm$0.04 & 1.41$\pm$0.37 & 62.5/74 &
 (3.7$\pm$0.1)$\times 10^{-9}$	& $26.61_{-2.58}^{+3.40}$ & $4.78_{-0.36}^{+0.34}$ & 0.81$\pm$0.31 & 70.0/74 \\
80102-04-82-00 & 2004-05-15 03:03-03:29 & (3.5$\pm$0.3)$\times 10^{-9}$
 & $>$170.1 & $1.53_{-0.07}^{+0.06}$ & $0.79_{-0.52}^{+0.53}$ & 53.5/74
 & (3.4$\pm$0.2)$\times 10^{-9}$ & $30.46_{-5.82}^{+13.12}$ & $4.25_{-0.83}^{+0.61}$ & $0.34_{-0.34}^{+0.45}$ & 57.4/74 \\
80102-04-83-00 & 2004-05-19 19:14-19:29 & $4.1_{-0.3}^{+0.5} \times
 10^{-9}$ 	& $>$180.6 & 1.48$\pm$0.05 & 1.29$\pm$0.45 & 59.5/74 &
 (3.9$\pm$0.2)$\times 10^{-9}$	& $26.66_{-3.22}^{+4.50}$ & $4.74_{-0.45}^{+0.42}$ & $0.86_{-0.37}^{+0.38}$ & 56.7/74 \\
80102-04-84-00 & 2004-05-23 11:00-11:12 & (3.9$\pm$0.4)$\times 10^{-9}$
 & $>$135.1 & $1.46_{-0.07}^{+0.06}$ & $1.34_{-0.51}^{+0.52}$ & 71.6/74
 & (3.7$\pm$0.2)$\times 10^{-9}$ & $25.29_{-3.86}^{+5.68}$ & $4.92_{-0.59}^{+0.55}$ & $0.92_{-0.45}^{+0.46}$ & 72.7/74 \\
80102-04-86-00 & 2004-05-29 10:15-10:39 & $3.8_{-0.3}^{+0.2} \times
 10^{-9}$ 	& $>$238.3 & $1.53_{-0.05}^{+0.02}$ &
 $1.01_{-0.36}^{+0.44}$ & 49.8/74 & (3.7$\pm$0.2)$\times 10^{-9}$ & $32.91_{-5.35}^{+10.85}$ & $4.08_{-0.65}^{+0.48}$ & $0.56_{-0.35}^{+0.36}$ & 56.0/74 \\
80102-04-87-00 & 2004-06-02 07:54-08:23 & (3.3$\pm$0.3)$\times 10^{-9}$
 & $>$154.3 & 1.52$\pm$0.06 & $1.11_{-0.23}^{+0.48}$ & 82.4/74 & (3.7$\pm$0.2)$\times 10^{-9}$	& $26.27_{-3.71}^{+5.56}$ & $4.62_{-0.53}^{+0.48}$ & 0.73$\pm$0.41 & 82.0/74 \\
80102-04-88-00 & 2004-06-06 06:19-06:45 & (3.7$\pm$0.3)$\times 10^{-9}$
 & $>$195.6 & $1.54_{-0.06}^{+0.05}$ & $1.11_{-0.44}^{+0.46}$ & 73.6/74
 & (3.1$\pm$0.2)$\times 10^{-9}$ & $28.41_{-4.11}^{+6.49}$ & $4.40_{-0.54}^{+0.47}$ & 0.77$\pm$0.39 & 72.0/74 \\
80102-04-89-00 & 2004-06-09 11:40-12:18 & ($4.2_{-0.3}^{+0.2}$)$\times 10^{-9}$ & -- & $1.54_{-0.05}^{+0.02}$ & $0.90_{-0.37}^{+0.41}$ & 63.0/74 & (4.0$\pm$0.2)$\times 10^{-9}$ & $31.03_{-4.68}^{+8.09}$ & $4.19_{-0.56}^{+0.46}$ & 0.52$\pm$0.36 & 61.9/74 \\
80102-04-90-00 & 2004-06-12 11:41-11:55 & (4.1$\pm$0.3)$\times 10^{-9}$ & $>$168.2 & $1.55_{-0.06}^{+0.05}$ & 1.00$\pm$0.46 & 88.4/74 & (4.0$\pm$0.2)$\times 10^{-9}$ & $27.41_{-4.34}^{+7.22}$ & $4.42_{-0.60}^{+0.51}$ & $0.63_{-0.39}^{+0.40}$ & 89.7/74 \\
80102-04-91-00 & 2004-06-15 23:08-23:30 & $4.0_{-0.3}^{+0.4} \times
 10^{-9}$ 	& $>$160.1 & 1.53$\pm$0.06 & 0.98$\pm$0.46 & 65.1/74 &
 (3.9$\pm$0.2)$\times 10^{-9}$	& $27.05_{-3.53}^{+5.09}$ & $4.50_{-0.47}^{+0.43}$ & $0.57_{-0.39}^{+0.40}$ & 61.5/74 \\
80102-04-92-00 & 2004-06-19 06:04-06:27 & (4.0$\pm$0.3)$\times 10^{-9}$
 & $184.8_{-57.9}^{+135.2}$ & 1.48$\pm$0.05 & 1.21$\pm$0.43 & 63.8/74 &
 (3.8$\pm$0.2)$\times 10^{-9}$	& $22.80_{-2.48}^{+3.23}$ & $5.07_{-0.42}^{+0.40}$ & $0.77_{-0.37}^{+0.38}$ & 58.2/74 \\
80102-04-93-00 & 2004-06-23 12:16-12:30 & $4.2_{-0.3}^{+0.2} \times
 10^{-9}$ 	& $>$243.1 & $1.53_{-0.05}^{+0.03}$ &
 $1.21_{-0.36}^{+0.44}$ & 66.1/74 & (3.9$\pm$0.2)$\times 10^{-9}$ & $28.58_{-3.58}^{+5.42}$ & $4.43_{-0.47}^{+0.41}$ & $0.87_{-0.36}^{+0.37}$ & 63.1/74 \\
80102-04-94-00 & 2004-06-28 02:28-02:48 & (3.8$\pm$0.3)$\times 10^{-9}$
 & $189.0_{-59.6}^{+141.6}$ & 1.50$\pm$0.05 & 0.94$\pm$0.43 & 51.3/74 &
 (3.6$\pm$0.2)$\times 10^{-9}$	& $22.95_{-2.57}^{+3.38}$ & $4.96_{-0.43}^{+0.41}$ & 0.50$\pm$0.37 & 46.8/74 \\
80102-04-95-00 & 2004-07-01 07:37-07:55 & (4.3$\pm$0.3)$\times 10^{-9}$
 & $>$167 & 1.51$\pm$0.05 & $1.22_{-0.40}^{+0.41}$ & 58.8/74 &
 (4.1$\pm$0.2)$\times 10^{-9}$ & $27.12_{-3.20}^{+4.48}$ & $4.52_{-0.43}^{+0.39}$ & 0.70$\pm$0.34 & 64.1/74 \\
80102-04-96-00 & 2004-07-03 20:01-20:14 & (4.8$\pm$0.4)$\times 10^{-9}$ & $>$174.9 & 1.51$\pm$0.05 & $1.32_{-0.42}^{+0.43}$ & 61.3/74 & (3.9$\pm$0.2)$\times 10^{-9}$ & $26.54_{-3.13}^{+4.34}$ & $4.63_{-0.43}^{+0.40}$ & 0.88$\pm$0.36 & 57.8/74 \\
90418-01-01-04 & 2004-07-10 03:27-03:42 & $5.9_{-0.5}^{+0.6} \times
 10^{-9}$	& $208.0_{-73.3}^{+211.1}$ & 1.53$\pm$0.06 &
 1.29$\pm$0.50 & 74.2/74 & (5.7$\pm$0.3)$\times 10^{-9}$ & $26.12_{-3.46}^{+4.99}$ & $4.52_{-0.49}^{+0.45}$ & 0.80$\pm$0.43 & 75.0/74 \\
90418-01-01-00 & 2004-07-10 05:00-05:17 & (5.8$\pm$0.5)$\times 10^{-9}$
 & $150.4_{-43.7}^{+90.0}$ & 1.50$\pm$0.06 & $1.76_{-0.48}^{+0.49}$ &
 59.3/74 & (5.6$\pm$0.3)$\times 10^{-9}$ & $23.57_{-2.74}^{+3.60}$ & $4.81_{-0.44}^{+0.42}$ & 1.22$\pm$0.42 & 59.7/74 \\
90418-01-01-01 & 2004-07-11 02:12-03:43 & (6.1$\pm$0.3)$\times 10^{-9}$
 & $164.7_{-23.3}^{+31.5}$ & 1.51$\pm$0.03 & 1.69$\pm$0.25 & 91.6/74 &
 $5.8_{-0.1}^{+0.2} \times 10^{-9}$ & $25.42_{-1.24}^{+1.41}$ & 4.49$\pm$0.17 & $0.86_{-0.20}^{+0.21}$ & 83.7/74 \\
90418-01-01-02 & 2004-07-12 03:22-04:36 & (6.4$\pm$0.3)$\times 10^{-9}$
 & $180.3_{-26.7}^{+37.0}$ & 1.51$\pm$0.03 & 1.74$\pm$0.25 & 80.5/74 &
 (6.0$\pm$0.2)$\times 10^{-9}$	& $26.06_{-1.25}^{+1.41}$ & $4.46_{-0.17}^{+0.16}$ & 0.94$\pm$0.20 & 53.5/74 \\
60705-01-65-00 & 2004-07-12 06:58-07:13 & (6.4$\pm$0.4)$\times 10^{-9}$ & $254.6_{-70.8}^{+147.2}$ &
 1.56$\pm$0.04 & 1.24$\pm$0.35 & 41.7/74 & (6.2$\pm$0.2)$\times 10^{-9}$ & $28.42_{-2.52}^{+3.29}$ & $4.18_{-0.30}^{+0.28}$ & 0.70$\pm$0.28 & 41.2/74 \\
90418-01-01-03 & 2004-07-13 02:59-04:04 & (6.7$\pm$0.3)$\times 10^{-9}$
 & $157.2_{-20.7}^{+27.4}$ & 1.51$\pm$0.03 & 1.78$\pm$0.25 & 49.9/74 &
 (6.3$\pm$0.2)$\times 10^{-9}$	& $25.65_{-1.24}^{+1.39}$ & 4.44$\pm$0.16 & 0.88$\pm$0.20 & 64.6/74 \\
60705-01-65-01 & 2004-07-15 07:45-08:24 & (7.0$\pm$0.5)$\times 10^{-9}$
 & $213.0_{-61.8}^{+134.9}$ & $1.56_{-0.05}^{+0.04}$ & 1.43$\pm$0.37 &
 66.0/74 & (6.7$\pm$0.3)$\times 10^{-9}$ & $25.57_{-2.54}^{+3.33}$ & $4.48_{-0.35}^{+0.33}$ & 1.00$\pm$0.32 & 62.2/74 \\
60705-01-66-00 & 2004-07-18 09:49-10:34 & $7.4_{-0.4}^{+0.5} \times
 10^{-9}$ & $200.6_{-48.4}^{+86.7}$ & 1.56$\pm$0.04 &
 $1.39_{-0.34}^{+0.34}$ & 50.2/74 & (7.1$\pm$0.3)$\times 10^{-9}$ & $28.18_{-2.77}^{+3.66}$ & $4.14_{-0.32}^{+0.30}$ & $0.72_{-0.28}^{+0.28}$ & 73.5/74 \\
60705-01-66-02 & 2004-07-21 21:42-22:00 & $7.8_{-0.4}^{+0.5} \times
 10^{-9}$  & $164.3_{-31.9}^{+49.4}$ & 1.54$\pm$0.04 &
 $1.60_{-0.32}^{+0.32}$ & 75.8/74 & (7.4$\pm$0.3)$\times 10^{-9}$ & $24.63_{-1.79}^{+2.13}$ & $4.52_{-0.25}^{+0.24}$ & $0.95_{-0.26}^{+0.27}$ & 79.2/74 \\
60705-01-67-01 & 2004-07-29 07:02-07:42 & (8.7$\pm$0.5)$\times 10^{-9}$
 & $167.6_{-30.0}^{+44.3}$ & 1.55$\pm$0.03 &
 $1.71_{-0.30}^{+0.30}$ & 69.1/74 & (8.3$\pm$0.3)$\times 10^{-9}$ & $25.36_{-1.69}^{+2.03}$ & $4.39_{-0.23}^{+0.22}$ & $0.99_{-0.24}^{+0.24}$ & 78.7/74 \\
60705-01-68-00 & 2004-08-01 02:36-03:18 & (8.7$\pm$0.5)$\times 10^{-9}$
 & $127.8_{-19.2}^{+26.6}$ & 1.54$\pm$0.04 &
 $1.85_{-0.30}^{+0.30}$ & 65.1/74 & (8.4$\pm$0.3)$\times 10^{-9}$ & $23.50_{-1.52}^{+1.77}$ & $4.50_{-0.23}^{+0.22}$ & $1.05_{-0.25}^{+0.25}$ & 86.4/74 \\
91105-04-13-00 & 2005-04-17 17:33-18:18 & (3.2$\pm$0.3)$\times 10^{-9}$ & $>$158.9 & 1.65$\pm$0.06 &
 1.61$\pm$0.47 & 81.1/74 & (3.0$\pm$0.2)$\times 10^{-9}$ & $24.33_{-3.63}^{+5.44}$ & $4.41_{-0.53}^{+0.49}$ & 1.51$\pm$0.42 & 71.4/74 \\ 
90704-01-12-00 & 2005-04-18 22:49-23:06 & $3.0_{-0.2}^{+0.3} \times 10^{-9}$ 	& $>$145.6 & 1.62$\pm$0.06 & $1.28_{-0.68}^{+0.46}$ &
 65.5/74 & $2.9_{-0.1}^{+0.2} \times 10^{-9}$	& $25.63_{-3.75}^{+5.80}$ & $4.32_{-0.53}^{+0.46}$ & 1.02$\pm$0.41 & 62.0/74 \\
91105-04-14-00 & 2005-04-20 22:39-23:28 & (2.7$\pm$0.2)$\times 10^{-9}$ & -- & $1.63_{-0.06}^{+0.04}$ & $1.16_{-0.45}^{+0.49}$ & 81.8/74 & ($2.7_{-0.1}^{+0.2}$)$\times 10^{-9}$ & $30.43_{-6.19}^{+15.98}$ & $3.96_{-0.90}^{+0.61}$ & $0.91_{-0.42}^{+0.43}$ & 83.2/74 \\
90704-01-13-00 & 2005-04-22 11:57-12:37 & (2.5$\pm$0.2)$\times 10^{-9}$
 & $>$149.9 & 1.54$\pm$0.06 & 1.27$\pm$0.50 & 77.4/74 & (2.4$\pm$0.1)$\times 10^{-9}$	& $25.39_{-4.33}^{+7.03}$ & $4.66_{-0.65}^{+0.58}$ & 0.97$\pm$0.44 & 75.4/74 \\
91105-04-15-00 & 2005-04-23 09:53-10:15 & (2.4$\pm$0.2)$\times 10^{-9}$ & -- & 1.57$\pm$0.04 &
 $1.05_{-0.23}^{+0.46}$ & 58.7/74 & ($2.4_{-0.1}^{+0.2}$)$\times 10^{-9}$ & $>$26.60 &
 $3.82_{-1.89}^{+0.74}$ & 0.76$\pm$0.44 & 56.1/74 \\ 
90704-01-13-01 & 2005-04-24 16:11-16:37 & (2.2$\pm$0.2)$\times 10^{-9}$
 & -- & $1.54_{-0.07}^{+0.04}$ & $1.30_{-0.45}^{+0.53}$ & 42.3/74 &
 (2.1$\pm$0.1)$\times 10^{-9}$ & $28.74_{-5.23}^{+9.78}$ & $4.45_{-0.73}^{+0.59}$ & 1.06$\pm$0.46 & 40.4/74 \\
90704-01-13-02 & 2005-04-27 08:09-08:42 & $1.7_{-0.1}^{+0.2}$ $\times 10^{-9}$ & -- & 1.54$\pm$0.05 &
 1.22$\pm$0.57 & 59.7/74 & $1.9_{-0.2}^{+0.1}$ $\times 10^{-9}$ & -- & $2.68_{-0.64}^{+1.32}$ & 0.99$\pm$0.54 & 54.4/74 \\ 
91105-04-16-00 & 2005-04-26 01:28-01:39 & $1.7_{-0.2}^{+0.1}$ $\times 10^{-9}$ & -- & $1.59_{-0.10}^{+0.05}$ &
 $0.63_{-0.56}^{+0.97}$ & 62.8/74 & (1.8$\pm$0.2)$\times 10^{-9}$ & $>$23.32 & $4.07_{-2.07}^{+0.91}$ & $<$1.35 & 59.7/74 \\
\hline
\label{table_5}
\end{tabular}
\end{center}
\end{minipage}}
\end{table*}

\end{document}